\documentstyle[emulateapj,psfig,apjfonts]{article}

\received{}
\accepted{}
\journalid{}{}
\articleid{}{}
 
\slugcomment{Accepted by the Astrophysical Journal}

\def\etal{{\rm et~al.\ }}
\def\HI{H\,{\sc i}}
\def\HII{H\,{\sc ii}}

\lefthead{GAENSLER ET AL}
\righthead{RADIO POLARIZATION FROM THE INNER GALAXY AT ARCMINUTE RESOLUTION}
    
\begin{document}
\title{Radio polarization from the inner Galaxy at arcminute resolution}
\author{B. M. Gaensler\altaffilmark{1,6}, 
John M. Dickey\altaffilmark{2}, 
N. M. McClure-Griffiths\altaffilmark{2},
A. J. Green\altaffilmark{3},  \\
M. H. Wieringa\altaffilmark{4} and R. F. Haynes\altaffilmark{5}}

\altaffiltext{1}{Center for Space Research, Massachusetts Institute of
Technology, 70 Vassar Street, Cambridge, MA 02139; bmg@space.mit.edu}
\altaffiltext{2}{Department of Astronomy, University of Minnesota, 116 Church Street SE,
Minneapolis, MN 55455}
\altaffiltext{3}{Astrophysics Department, School of Physics, University of Sydney, NSW 2006, Australia}
\altaffiltext{4}{Australia Telescope National Facility, CSIRO, Locked Bag 194, Narrabri, NSW 2390, Australia}
\altaffiltext{5}{Australia Telescope National Facility, CSIRO, PO Box 76, Epping, NSW 1718, Australia}
\altaffiltext{6}{Hubble Fellow}

\begin{abstract}

The Southern Galactic Plane Survey (SGPS) is an \HI\ and 1.4-GHz continuum
survey of the 4th quadrant of the Galaxy at arcmin resolution.  We
present here results on linearly polarized continuum emission from an
initial 28-square-degree Test Region for the SGPS, consisting of 190
mosaiced pointings of the Australia Telescope Compact Array, and
covering the range $325\fdg5 < l < 332\fdg5$, $-0\fdg5 < b < +3\fdg5$.
Complicated extended structure is seen in linear polarization
throughout the Test Region, almost all of which has no correlation with
total intensity. We interpret the brightest regions of polarized
emission as representing intrinsic structure in extended polarization,
most likely originating in the Crux spiral arm at a distance of
3.5~kpc; fainter polarized structure is imposed by Faraday rotation in
foreground material.  Two large areas in the field are devoid of
polarization. We argue that these voids are produced by foreground
\HII\ regions in which the magnetic field is disordered on scales of
$\sim0.1-0.2$~pc.  We also identify a depolarized halo around the
\HII\ region RCW~94, which we suggest results from the interaction
of the \HII\ region with a surrounding molecular cloud.

\end{abstract}


\section{Introduction}
\label{sec_intro}

Soon after linear polarization was first detected from celestial radio
sources, it became apparent that the entire Galactic Plane was
a significant source of polarized emission (\cite{sw61}).
There appear to be at least two components to this emission
(e.g.\ \cite{jfr87}; \cite{dhjs97}): polarization
from discrete supernova remnants (SNRs),
and a diffuse polarized background produced by the interaction
of the relativistic component of the interstellar medium (ISM)
with the Galactic magnetic field.

In the presence of magnetic fields and free electrons,
electromagnetic radiation propagates in two
circularly-polarized orthogonal modes of different velocities.
Linearly polarized radiation passing through a magneto-ionized
medium will thus emerge with its position angle
rotated, an effect known as {\em Faraday rotation} (\cite{far44}; \cite{cp61}).
If a source emits linear polarization with an intrinsic position angle 
$\Theta_0$, then the measured position angle $\Theta$ 
at an observing wavelength 
$\lambda$ is given by:
\begin{equation}
\Theta = \Theta_0 + {\rm RM}~\lambda^2.
\label{eqn_rm}
\end{equation}
In this expression, the {\em rotation measure}\ (RM), in units
of rad~m$^{-2}$, is defined by :
\begin{equation}
{\rm RM}~ = K \int B \cos \theta~n_e dl,
\end{equation}
where $K = 0.81$~rad~m$^{-2}$~pc$^{-1}$~cm$^3$~$\mu$G$^{-1}$,
$B$, $\theta$ and $n_e$ are the magnetic field strength, inclination
of the magnetic field to the line of sight and electron
density respectively, and
the integral is along the line of sight
from the observer to the source.

Faraday rotation can be a powerful probe of 
Galactic sources, and has been used to determine
the magneto-ionic properties of both
individual objects (e.g.\ \cite{hc80}; \cite{mldg84}) and of the Galaxy
as a whole (e.g.\ \cite{sk80}; \cite{hq94}).
Extragalactic sources (\cite{ccsk92}),
pulsars (\cite{hmq99}) and  extended polarized
emission from the entire Galaxy (\cite{spo84}) have all been used
as background sources against which Faraday rotation can
be studied. However, with a single-dish telescope all 
such studies have their limitations --- studies
of extragalactic sources suffer from
confusion in the Galactic Plane, pulsars are of too low
spatial density to study specific regions of the sky in detail,
and the spatial resolution with which extended emission
can be imaged is generally poor.

In more recent years, instruments and techniques have been developed
which allow interferometers to map large regions of the sky in
polarization. Consequent polarimetric surveys with both the Westerbork
and DRAO Synthesis Telescopes do not suffer from any of the problems
described above for single-dish observations, and can thus study
polarization and Faraday rotation at high spatial resolution. These
studies have revealed several unusual features
with no counterparts either in total intensity or at any other wavelength
(\cite{wdj+93};  \cite{gldt98}; \cite{gld+99}; \cite{hkd00}).  These
results demonstrate that there are components of the ISM which can only
be studied through polarimetry.

The above-mentioned surveys have been carried out at high Galactic
latitudes and/or in the 2nd Galactic quadrant, and thus provide
important information on the magneto-ionic properties of local gas and
of the outer Galaxy.  However, it is also of considerable interest to
study the polarimetric properties of the inner Galaxy, in which there
are undoubtedly regions of stronger magnetic fields, higher densities
and increased levels of turbulence. The complexity of polarization in
this part of the sky has been demonstrated by the spectacular
single-dish polarization surveys of Duncan
\etal (1997a\nocite{dhjs97}, 1999\nocite{drrf99}), but a full
understanding of the emission in these regions requires
the higher angular
resolution which only an interferometer can provide.

The Australia Telescope Compact Array (ATCA) is ideally suited to
such a study. Its antipodean location allows it to observe
almost the entire 4th Galactic quadrant, the high polarimetric purity
of its receivers ensures that polarized signals are uncontaminated
by leakage from total intensity, its mosaicing capability allows
it to easily map large parts of the sky, its multi-channel
continuum mode allows accurate studies of Faraday rotation,
and its compact antenna configurations
give it good sensitivity to extended structure. 
Motivated by these capabilities, we are in the
process of mapping 
the entire Galactic Plane south of
declination $\sim-35^\circ$ in 1.4-GHz polarization;
observations are being made simultaneously in the \HI\ line.
Together with single-dish \HI\ data being taken
with the 64-m Parkes radiotelescope (\cite{mdg+00a}), all these data form the {\em
Southern Galactic Plane Survey}\ (SGPS; \cite{dmg+99}; 
\cite{mdg+00c})\footnote{see also the SGPS WWW page, at
{\tt http://www.astro.umn.edu/\~{ }naomi/sgps.html .}},
the ATCA component of
which will cover the
Galactic coordinates $253^\circ < l < 358^\circ$, $-1^\circ < b < 1^\circ$.
The polarimetric component of this data-set will have a spatial
resolution of $\sim$1~arcmin, and a theoretical sensitivity of
$\sim$0.25~mJy (where 1~jansky~$= 10^{-26}$~W~m$^{-2}$~Hz$^{-1}$).

Before embarking on the full SGPS, we carried out ATCA
observations on a small Test Region, covering the range
$325\fdg5 < l < 332\fdg5$, $-0\fdg5 < b < +3\fdg5$.
A preliminary image of this region, at a resolution
of $\sim$5~arcmin, was presented by Dickey (1997\nocite{dic97}).
We present here our entire data-set on polarization
from the SGPS Test Region, at the final survey resolution
of 1~arcmin. In Section~\ref{sec_obs} we describe the
aspects of SGPS observations and data-reduction relevant
to the polarimetric component of the survey, while
in Section~\ref{sec_results} we present the distribution
of polarized emission and of rotation measure which we
have determined from these data. In Section~\ref{sec_discuss},
we interpret these results in terms of
intrinsic polarization and Faraday rotation in the Galactic Plane.
The present paper focuses on extended regions of linear polarization;
in a separate paper by Dickey \etal (in preparation), we show how
the RMs measured towards point sources in the SGPS Test Region
can constrain models of the Galactic magnetic field.

\section{Observations and Reduction}
\label{sec_obs}

Observations of the SGPS Test Region
were made with the Australia Telescope Compact
Array (ATCA; \cite{fbw92}), a synthesis telescope
located near Narrabri, NSW, Australia. Details of
the observing dates, array configurations and pointing
centers for observations of the Test Region
will be described in a forthcoming
paper by McClure-Griffiths \etal (2000d\nocite{mgd+00}).
All reduction was carried out in the {\tt MIRIAD}\ package
(\cite{sk98}),
using standard techniques unless otherwise specified.

Each antenna of the ATCA receives incident radiation in two orthogonal,
linearly polarized modes, $X$ and $Y$. Four polarization spectra,
$XX$, $YY$, $XY$ and $YX$, are consequently measured for each correlation.
These data are recorded as 32 4-MHz channels
across a 128-MHz bandwidth, with a center frequency of 1384 MHz.  The
ATCA's continuum mode includes an inherent triangular weighting for the
lag-spectrum, meaning that alternate channels are redundant; we thus
discarded every second channel. We also discarded four channels on either
edge of the bandpass because of low signal-to-noise, and removed
channels centered on 1376, 1384 and 1408 MHz because of interference
(both external and self-generated). The resulting data consisted
of nine spectral channels across 96 MHz of bandwidth.
Additional editing was then carried out to
remove time-ranges and visibilities corrupted by instrumental problems 
and radio interference.

Flux density and bandpass calibrations were applied to the data using a
$\sim$10-min observation, made once each synthesis, of PKS~B1934--638,
for which we assumed a flux density of 14.94~Jy
at 1384~MHz (\cite{rey94}). Variations
of the atmospheric gain above each antenna as a function of time were
determined using a 3-min observation of either MRC~B1438--481 or
MRC~B1613--586 every $\sim$100 minutes.  Observations of these
calibrators over a wide range in parallactic angle were used also to
solve for the instrumental polarization characteristics of each antenna
(\cite{shb96}).  The calibrated data were then converted to the four
standard Stokes parameters, $I$, $Q$, $U$ and $V$.

The visibilities from all observations were then inverted to produce
images of the Test Region using all baselines between 31~m and 765~m,
and employing a uniform weighting scheme to minimize sidelobe
contributions from strong sources.  
A mosaiced image of the entire Test Region was formed
for each of the nine spectral channels and four Stokes parameters
(36 images in total) by linearly combining data from
the 190 separate pointings (\cite{chu93}; \cite{ssb96}).
The four Stokes images for each spectral channel were
deconvolved jointly using the maximum entropy algorithm {\tt PMOSMEM}\
(\cite{sbd99}). This approach successfully recovers the large-scale
structure measured by the mosaicing process (\cite{er78}), and also
uses the ``maximum emptiness'' criterion, which allows the Stokes $Q$,
$U$ and $V$ maps to take on negative values.  No constraints were given
to the deconvolution process regarding the total flux density in each
image or sub-region to be deconvolved.  Following deconvolution, each
of the 36 images was smoothed with a gaussian restoring beam of FWHM
$87''\times67''$, the major axis being oriented approximately
north-south.  The images were then re-gridded into Galactic
coordinates using a gnomonic (``TAN'') projection (see 
\cite{gc95} and references therein).  
Images of linearly polarized intensity, $L = (Q^2 +
U^2)^{1/2}$, linearly polarized position angle, $\Theta =
\frac{1}{2} \tan^{-1}(U/Q)$, and uncertainty
in position angle, $\Delta\Theta = \sigma_{Q, U}/2L$,
were then formed from each pair of $Q$ and
$U$ images, where $\sigma_{Q, U}$ is the RMS sensitivity
in the $Q$ and $U$ images.
A correction was applied to the $L$ images to account for
the Ricean bias produced when $Q$ and $U$ are combined (\cite{kbe86}),
The nine $L$ maps (one per spectral channel) were then averaged
together to make a final image of $L$ for the entire Test Region.
The position angle maps were also corrected so that position
angles were relative to Galactic, rather than equatorial, coordinates.

Note that an alternative approach to the data reduction would have been
to form a single image for each Stokes parameter using multi-frequency
synthesis, whereby the data from all spectral channels are imaged
simultaneously (\cite{sw94}).  While this approach results in improved
$u-v$ coverage and image fidelity, the resulting images of polarized
intensity would suffer from significant depolarization wherever Faraday
rotation across the 96-MHz observing bandwidth was significant. By
combining information from each channel only after we have formed
separate images of $L$ for each, we reduce the effects of bandwidth
depolarization to only what is incurred across each 8-MHz channel.

\section{Results}
\label{sec_results}

\subsection{Images of Polarized Intensity}

Over most of the field, data at a given position contains contributions
from nine adjacent pointings; the resultant theoretical sensitivity in
each Stokes parameter is $\sim$0.5~mJy~beam$^{-1}$. In a strip of width
$0\fdg5$ around the edge of the field, the sensitivity has a higher value
of $\sim$0.9~mJy~beam$^{-1}$ because the number of adjacent fields is
smaller.

A total intensity image is shown in Fig~\ref{fig_ext_I}.  This image,
which will be discussed in more detail by McClure-Griffiths \etal 
(2000d\cite{mgd+00}), shows a variety of SNRs and \HII\ regions along the
Galactic Plane, embedded in more diffuse emission to which the
available interferometric spacings are only partially sensitive.
Further from the Plane ($b \ga 1^\circ$), only unresolved, presumably
background sources are seen.  The structure seen in this image is very
similar to that seen at 843~MHz and at comparable resolution in the
Molonglo Galactic Plane Survey (Fig~4 of \cite{gcl98}).  For comparison
purposes, in Fig~\ref{fig_I_11cm} we show the total intensity from the
2.4~GHz survey of Duncan \etal (1995\nocite{dshj95}), which was carried
out with the 64-m Parkes radiotelescope at an angular resolution of
$10\farcm4$. Although the single-dish observations detect a
considerable amount of extended emission to which the interferometer is
not sensitive, the same main features are identifiable in each of
Figs~\ref{fig_ext_I} and \ref{fig_I_11cm}.

Linearly polarized emission from the SGPS data-set is shown in
Figs~\ref{fig_ext_L}, \ref{fig_ext_L_2} and \ref{fig_ext_theta}.
Figs~\ref{fig_ext_L} and \ref{fig_ext_L_2} display
the intensity of extended linear polarization using two different
greyscale ranges, while Fig~\ref{fig_ext_theta} illustrates
the polarization position angle (averaged across the nine
spectral channels) for these data. 

Five distinct classes of feature can be seen in the ATCA images:
\begin{enumerate}

\item Structures in polarization and depolarization corresponding
directly to sources seen in total intensity --- these features are
discussed in more detail in Section~\ref{results_discrete}.

\item Diffuse polarization across the entire field of view.
Fig~\ref{fig_ext_theta} demonstrates that
this emission is composed of discrete patches, in each
of which the position angle of polarization
is approximately uniform. However, sudden changes in
position angle are seen between one patch and the next.

\item Narrow ``canals'' of greatly reduced polarization running
randomly throughout the diffuse component, best seen
in Fig~\ref{fig_ext_L_2}.

\item Large ``voids'' of reduced polarization, several degrees across,
most clearly seen in  Fig~\ref{fig_ext_L_2}.
There are two particularly distinct such voids, the outer
perimeters of which we have approximated by
ellipses in Fig~\ref{fig_ext_L_2}: the 
first of these (henceforth referred to as ``void 1'')
is approximately circular, centered on approximately 
$(l, b) = (332\fdg4, 
+1\fdg4)$\footnote{Throughout this paper, we refer to the 
position of a source
with coordinates $l = l_0$, $b=b_0$ using the notation $(l_0, b_0)$.}
with a radius $1\fdg0-1\fdg5$. The second void (``void 2'') 
runs along the bottom edge of the field, with center
$(328\fdg2, -0\fdg5)$ and minor-axes in the $l$ and
$b$ directions of $\sim2^\circ$ and $\sim1\fdg5$ respectively.
Neither void has any counterpart in total intensity, when
viewed either with the interferometer (Fig~\ref{fig_ext_I})
or with a single-dish (Fig~\ref{fig_I_11cm}).

\item Polarized point sources throughout the field of view,
best seen in Fig~\ref{fig_ext_L}.

\end{enumerate}

In Fig~\ref{fig_L_11cm} 
we show the linearly polarized intensity from
the 2.4~GHz Parkes survey (\cite{dhjs97}) at a resolution of
$10\farcm4$, covering the same area as the SGPS
Test Region.
Some significant similarities and differences can
immediately be seen by comparing 
Figs~\ref{fig_ext_L} and \ref{fig_ext_L_2} with Fig~\ref{fig_L_11cm}.

First, we note that the two voids described above are clearly seen
also at 2.4~GHz. In the wider field images presented in
Fig~4 of Duncan \etal (1997a\nocite{dhjs97}), the two voids
can be seen to be nearly complete shells.

Duncan \etal (1997a\nocite{dhjs97}) identified two components
in the polarization they detected: bright, extended regions of
emission, and fainter regions with a patchy or clumpy morphology.
Both components can clearly be seen in Fig~\ref{fig_L_11cm}. 
The bright emission extends
approximately horizontally across the range $327^\circ < l <
331^\circ$, with a prominent spur extending away from the
plane at $l=329^\circ$. Fainter, patchier emission can be seen over 
almost all the rest of the field.
Generally speaking, most of the bright and extended emission
seen at 2.4~GHz is well-correlated with polarized emission
seen in our 1.4~GHz data. On the other hand, the fainter
patches seen at 2.4~GHz show little correspondence with polarization
at 1.4~GHz.

One notable exception to this is in the range $328\fdg5 < l <
329\fdg5$, $2\fdg0 < b < 3\fdg5$, where the boundaries of the bright
spur seen at 2.4~GHz seem to correspond to a region of low
polarization at 1.4~GHz.  Meanwhile, the hole in polarization seen just
to the right of the 2.4~GHz spur corresponds to bright polarization at
1.4~GHz.

In circular polarization (not shown here), instrumental artifacts can be
seen at the positions of the brightest continuum sources,
corresponding to leakage from Stokes~$I$ into
Stokes~$V$.  The only other source showing
emission at more than a level of 5$\sigma$ in
both Stokes $I$ and $V$ is the pulsar PSR~B1557--50
at position $(330\fdg69, +1\fdg63)$,
which has $I = 15$~mJy and $V=-5$~mJy.\footnote{Circular
polarization is defined such that $V>0$ corresponds
to right-hand circular polarization.}
This is in reasonable agreement with the previously
measured value of $V/I \approx -0.15$ for this source
(\cite{hmxq98} and references therein).

\subsection{Determination of Rotation Measures}
\label{sec_res_rm}

At almost every point where linear polarization is detected, the
position angle of this polarization varies from channel to channel
across the observing band as a result of Faraday rotation. This
effect can be used to calculate the rotation measure (RM) of
the polarization. 

We calculate the RM as a function of position from images
of $\Theta$ and $\Delta\Theta$ using the
{\tt MIRIAD}\ task {\tt IMRM} (\cite{sk98}). The algorithm used
by {\tt IMRM}\ is as follows (the nine channels being
considered in order of decreasing frequency): 

\begin{enumerate}

\item For each pixel, a function of the form 
given by Equation~(\ref{eqn_rm}) is fitted
to the values of $\Theta$, $\Delta\Theta$ and $\lambda$
corresponding to the first two channels only.
This produces an initial estimate of the values of
RM and $\Theta_0$. 

\item These estimates are then
used to predict
values of $\Theta$ at the other seven frequencies.

\item Integer multiples of $\pi$ radians are then added
or subtracted to the position angles at the other
frequencies in order to make their position
angles as close as possible to the predicted values
of $\Theta$. 

\item A least squares fit is then used
to solve for RM and $\Theta_0$ using the data
from all nine spectral channels.

\item This RM is subtracted from the data,
and steps 1 through 4 are repeated.

\item The RMs determined in the two executions
of step 4 are added together to give a final estimate of the RM.

\end{enumerate}

We only carried out this computation at pixels for which the
polarized intensity met the criterion $L > 5 \sigma \approx
3$~mJy~beam$^{-1}$, for which the reduced $\chi^2$ of
the least squares fit in Step 3 above satisfied $\chi^2_r < 1$,
and which did not correspond to polarized point sources
(which are treated separately in Section~\ref{results_compact} below).
Using this approach, we were able to calculate
a RM for 13\% of the $\sim9.9\times10^5$ pixels in the map;
the resulting distribution of RM is shown in Fig~\ref{fig_ext_rm}.
The uncertainty in an individual RM measurement
depends on signal-to-noise, but is typically in the range $\Delta$RM~$=
20-40$~rad~m$^{-2}$.  

It can be seen that the RMs are mostly small and negative ---
50\% of the RMs have magnitudes smaller
than $\pm$25~rad~m$^{-2}$ and 98\% are smaller than 
$\pm$100~rad~m$^{-2}$. 
The mean RM of the entire region is $-12.9\pm0.1$~rad~m$^{-2}$, 
with a median of $-12.7$~rad~m$^{-2}$; 
Some RMs of magnitudes of up to
$\pm400$~rad~m$^{-2}$ are seen --- these are predominantly on the edges
of voids, particularly in the region $331^\circ < l < 332^\circ$. In
Fig~\ref{fig_rm_plot_2} we show RMs towards four representative
regions. We have chosen two RMs of small magnitude and two of large
magnitude; it can be
seen that in all cases the fit to the data is very good,
as demanded by the constraint imposed on $\chi^2$.

\subsection{Discrete Structures in Polarization}
\label{results_discrete}

\subsubsection{Compact Sources}
\label{results_compact}

It is difficult to identify polarized emission from compact sources
in Figs~\ref{fig_ext_L} and \ref{fig_ext_L_2}, because of the
diffuse polarization which covers most of the field. To
look for polarization on small scales, we thus re-imaged
the data in $I$ and in $L$ as described in Section~\ref{sec_obs}, but
only using baselines between 210~m and 765~m. The resulting images
are sensitive only to scales smaller than $\sim3'$.

The RMS noise in the resulting $Q$ and $U$ images 
is $\sim0.55$~mJy~beam$^{-1}$, and in the
corresponding total intensity image is $\sim$0.8~mJy~beam$^{-1}$.
Adopting a 5$\sigma$ detection threshold, we consider
a source to be linearly polarized if
its surface brightnesses exceed $L >
2.75$~mJy~beam$^{-1}$ and $I > 4.0$~mJy~beam$^{-1}$.  We only
consider sources for which $L/I \ga 0.005$, since instrumental leakage
can introduce spurious polarization below this level. 

With these criteria, we find 21 compact sources showing linear
polarization. We determined the rotation measure (RM) of each source
using the method described in Section~\ref{sec_res_rm} above,
but relaxing the requirement that $\chi_r^2 < 1$.
The data and the RM function fitted
to them in each case are shown in Fig~\ref{fig_rm_plot}.
The 21 sources are listed in Table~\ref{tab_pol};
columns 1 through 7 indicate respectively the source
number, the Galactic longitude and latitude
of the pixel with the highest linearly polarized intensity,
the values of $L$ and $I$ at this pixel, the ratio $L/I$ for this pixel,
and the RM for this pixel.

These sources will not be discussed further here. In an associated
paper by Dickey \etal (in preparation), we show how the RMs which we
have determined for these sources can be applied to models for the
magnetic field structure of the Galaxy.

\subsubsection{Supernova Remnants}

There are seven SNRs from the catalogue of Green (2000\nocite{gre00})
in the Test Region. While polarization is detected in the direction
of all these sources, the presence of significant polarized emission
throughout the field makes it difficult to determine whether this
polarization is associated with a SNR or just happens to lie
along the same line of sight. Only in the case of G327.4+0.4~(Kes~27;
\cite{mck+89})
is the morphology of polarized emission correlated
with the total intensity from the SNR, and thus likely
to be associated with it. 

\subsubsection{\HII\ Regions}

There are $\sim$30 radio \HII\ regions within the SGPS Test Region
(e.g.\ \cite{ave97}).  Most of these sources are either near the edge
of the field, or are projected against void~2,
and so it is not possible to
determine whether they have any effect on the polarized emission at
their positions. However, there is one \HII\ region, RCW~94
(=~G326.3+0.8), whose presence is clearly seen in
polarization. 

The polarization in the direction of RCW~94 is shown
in Fig~\ref{fig_hii_pol}.
There is 
bright linearly polarized emission in the vicinity,
particularly to the lower-left of the \HII\ region
where the surface brightness in polarization is 5--10~mJy~beam$^{-1}$.
However, at positions coincident with the interior of the \HII\ region, the
polarized intensity is significantly reduced, falling in the range
3--5~mJy~beam$^{-1}$.  More notably, an annulus of width
$\sim5'$ almost completely surrounds RCW~94, in which the level of
polarization is reduced even further --- in most of this halo no
polarization is detected at all, down to a limit of
$\sim$2~mJy~beam$^{-1}$.  Both these effects are
demonstrated quantitatively in 
Fig~\ref{fig_rcw94_slice}, in which profiles of $I$ and $L$
are shown for the slice marked in Fig~\ref{fig_hii_pol}.
Both the region of reduced polarization in the interior of
RCW~94 and the very low levels of polarization around it
can clearly be seen.

Another \HII\ region,
G326.7+0.5 is 30~arcmin to the lower-left of
RCW~94.  Linear polarization in the interior of the source has a mean
surface brightness of $\sim$5~mJy~beam$^{-1}$, brighter than in the
interior of RCW~94 but still approximately half the brightness of
the surrounding emission. Although not as striking as for RCW~94, there
is possibly a partial halo of very low polarization surrounding
G326.7+0.5, most noticeable to the left of the source. This region
extends to approximately twice the radius of the \HII\ region.

\section{Discussion}
\label{sec_discuss}

\subsection{Interferometric Observations of Faraday Rotation}
\label{sec_disc_interf}

Before interpreting 
the diffuse polarized emission seen in the Test Region,
it is important to consider
what an interferometer detects when it observes polarized
emission which has undergone Faraday rotation.

In Appendix~\ref{appendix_1} we consider a simple model involving a
uniform polarized background, which passes through both a uniform
screen of RM $R_s$ and also a compact cloud of RM $R_c$. The background
polarization, after passing through the screen, has its position angle
changed,  but is still uniform in $I$, $Q$ and $U$. Since the
interferometer has no sensitivity to a constant offset in intensity,
this component of the emission is not detected. However, the emission
passing through the compact cloud emerges with a different polarized
position angle from elsewhere. Thus while the emission in Stokes~$I$ is
unaffected by passage through the cloud and so remains invisible to the
interferometer, structure is produced in $Q$ and $U$ on scales to which
the interferometer is sensitive.

This polarized emission has two important properties
(see Appendix~\ref{appendix_1}):

\begin{enumerate}
\item A polarized source will be detected at the position
of the cloud which has no counterpart in total intensity.
The surface brightness of this polarized emission can be
up to twice that of the (undetected) uniform polarized background.

\item The measured RM at this position will be equal to $R_s + R_c/2$. Thus
even though the uniform screen produces no detectable structure,
its RM is still preserved in any emission detected. Furthermore,
the apparent RM of the cloud  is {\em half}\ its true RM.
\end{enumerate}

\subsection{Diffuse Polarized Emission}

Figs~\ref{fig_ext_L} and \ref{fig_ext_L_2} dramatically demonstrate
the presence of large amounts of diffuse polarized emission
throughout the SGPS Test Region. Comparing Figs~\ref{fig_ext_I}
and \ref{fig_ext_L}, it is clear that much of this linearly
polarized emission has no counterpart in total intensity.
This can be understood in terms of the limited range
of scales to which the interferometer is sensitive.
The largest scale which the telescope can detect is usually determined
by the shortest antenna spacing.  In this case that spacing is 31~m,
which corresponds to a spatial scale on the sky of $\sim25'$. However,
the mosaicing process allows us to recover information on larger scales
(\cite{cor88}; \cite{ssb96}), in this case up to $\sim35'$.  Thus if
there is diffuse polarized emission in the Test Region such that its
total intensity is smoothly distributed, but the position angles of
polarization vary on scales $<35'$, then $Q$ and $U$ will contain
structure small enough to be detected, but the emission in Stokes~$I$
will not be seen.  This phenomenon is a common occurrence, and is seen
whenever an interferometer surveys significant areas of the sky in
polarization (\cite{wdj+93}; \cite{dhr+98}; \cite{gld+99};
\cite{hkd00}).

\subsubsection{Origin of the Polarized Emission}
\label{sec_disc_origin}

There are two possible mechanisms which can produce structure in
polarized intensity when the total intensity is smooth.  The first
possibility is that considered in Section~\ref{sec_disc_interf} above,
in which the intrinsic polarized intensity is smooth,
but for which compact foreground structure
induces Faraday rotation on small scales.
This will then produce structure in $Q$ and $U$ on
smaller scales to which the interferometer is sensitive
(see detailed discussion by \cite{wdj+93}).
If this is occurring, we expect to see no correlation between
polarized and total intensity, since all the structure in polarization
is generated by Faraday rotation alone. 
We also expect to see little correlation between
structures observed at widely-separated observing frequencies, since 
the amount of Faraday rotation (and hence the level of
structure produced by the foreground screen) is
strongly frequency-dependent.

The second possibility is that the diffuse synchrotron emission has
intrinsic structure in polarization, produced by a non-uniform magnetic
field in the emitting regions. The brightest polarized structures in
the single-dish 2.4~GHz polarization survey of Duncan
\etal (1997a\nocite{dhjs97}) are correlated with features in total
intensity and have position angles across them which are particularly
uniform, leading those authors to argue that this polarization
corresponds to intrinsic structure in the emitting regions. 

For the data presented here, a strong case can be made that both of the
effects just discussed are contributing to the observed polarization.
Since effects resulting from Faraday rotation are strongly dependent both on
frequency and angular resolution (e.g.\ \cite{bur66}; \cite{tri91}),
the fact that most of the bright extended regions of polarization
seen at 2.4~GHz are also seen at 1.4~GHz argues strongly that these
structures must be intrinsic to the source of emission (as claimed by
Duncan \etal [1997a\nocite{dhjs97}] from the 2.4~GHz data alone).
In the particular case of the bright polarized emission seen
in Fig~\ref{fig_L_11cm} in the range
$327^\circ < l < 331^\circ$, Duncan \etal (1997a\nocite{dhjs97})
suggest a possible association with the
large SNR G325+0, the limbs of which can also faintly be seen in total
intensity in 2.4~GHz data (\cite{dshj97}). 

The one exception to the 1.4/2.4~GHz correspondence between regions of
bright polarization is the ``spur'' of polarization
seen near $(329^\circ ,~+3^\circ)$, which is also
thought to be part of SNR~G325+0 (\cite{dshj97}), but
for which emission seems
anti-correlated between 1.4~GHz and 2.4~GHz. This can be
understood if internal depolarization is occurring at 1.4~GHz,
in which polarized emission from different depths within
the source undergoes differing amounts of Faraday rotation,
and destructively interferes to produce reduced levels
of overall polarization. We define the
``depolarization factor'', $p_i$, to be the ratio
of detected to intrinsic intensity of linear polarization
after internal depolarization.
In the simplest models
for internal depolarization (\cite{bur66}; \cite{sbs+98}),
we find that:
\begin{equation}
p_i  = \left|\frac{\sin R\lambda^2}{R\lambda^2}\right|
\label{eqn_depol_internal}
\end{equation}
where $R$ is the rotation measure through the entire source. It can be seen
from Equation~(\ref{eqn_depol_internal}) that the
observed polarization properties  of this spur
($p_i \sim 0$ at 1.4~GHz, but $p_i \sim 1$ at 2.4~GHz)
can be accounted for if the internal RM within
the spur is of magnitude $|R| \sim70$~rad~m$^{-2}$,
a reasonable value if the spur is indeed part of a SNR
(e.g.\ \cite{mr94}; \cite{dm98}).

If we now consider regions of fainter polarized emission,
we find that the location of and structure within such regions
is poorly correlated
between 1.4 and 2.4~GHz. This cannot easily be explained
if this structure is intrinsic to the source of emission,
but is exactly as would be
expected if it is primarily produced by Faraday rotation in
a foreground screen.  There is generally more structure seen at 1.4~GHz
than at 2.4~GHz, a result which we expect to be due to three separate
effects. First, the angle through which a foreground screen will rotate
polarized light is $\sim3$ times smaller at 2.4~GHz than at 1.4~GHz, so
we expect to see much less patchy structure induced by the screen when
viewed at the higher frequency.  Second, if the position angle of
polarization varies on scales which are resolvable at 1.4~GHz but not
by the lower-resolution 2.4~GHz data, this can cause
beam-depolarization at the higher frequency (see
discussion in Section~\ref{sec_disc_void_model} below).
Finally, if there are
high RMs along the line-of-sight, this can then cause significant
bandwidth depolarization (see Equation~[\ref{eqn_depol_band}] below) at 2.4~GHz
(where 1000~rad~m$^{-2}$ causes $\sim100^\circ$ of rotation across the
145~MHz observing bandwidth) but not at 1.4~GHz (where the same RM
produces $\sim30^\circ$ across each 8~MHz spectral channel).

By comparison with the 2.4~GHz data,
we have argued above that structure in 1.4-GHz linear polarization which
is intrinsic to the emitting regions is generally brighter
than that produced by foreground Faraday rotation. 
In this case, we can then
constrain the excess RMs which produce the latter structures. From
Equation~(\ref{eqn_appendix_final}), the intensity of polarized 
emission produced via foreground Faraday
rotation is $P = 2P_0\sin(R_c \lambda^2)$, where $P_0$ is the intensity
of the background polarization and $R_c$ is the RM of the
foreground material. If we
set $P < P_0$, we find that $|R_c| < 10$~rad~m$^{-2}$ to
produce the observed structure, of which we will only detect 
a contribution of magnitude $|R_c/2|$
(from Equation~[\ref{eqn_appendix_final}]). Excess RMs of this magnitude are too small to be
discerned from our data, and we thus do not expect to see enhanced RMs
in regions where Faraday-induced polarization is occurring. Indeed,
examination of Fig~\ref{fig_ext_rm} shows no noticeable enhancements in
RM in any particular regions, except around the edges of voids.  We
note that in the most clear-cut case of Faraday-induced polarization
structure (\cite{wdj+93}), the excess RMs measured were indeed of magnitude
$\sim$5~rad~m$^{-2}$, just as predicted here.

\subsubsection{Distance to the Polarized Emission}
\label{sec_disc_dist}

We have just shown that the contribution to the RM from discrete
foreground clouds cannot be more than $\sim$5~rad~m$^{-2}$. Thus the
RMs seen in Fig~\ref{fig_ext_rm} must be due almost entirely to Faraday
rotation in the diffuse ISM between the source and the observer (i.e.\
the term $R_s$ in Appendix~\ref{appendix_1}).  We can estimate the
distance from us at which the polarized emission is being produced by
comparing the RMs for this diffuse emission to those measured for
pulsars in the same part of the sky. If we only consider pulsars in the
area $320^\circ < l < 340^\circ$ and $|b| < 3^\circ$, we find ten
measurements of pulsar RMs in the catalogue of Taylor \etal
(1995\nocite{tmlc95}), plus an additional eight more recent measurements
listed by Han, Manchester \& Qiao (1999\nocite{hmq99}).  Using distances for
these pulsars as determined from their dispersion measures
(\cite{tc93}), we find that the only pulsars in this sample for which
--100~rad~m$^{-2}$~$<$~RM~$<$0~rad~m$^{-2}$ are the six sources whose
distances lie in the range 1.3--4.5~kpc; pulsars at smaller distances have
RM~$>0$~rad~m$^{-2}$, while at larger distances we generally find that
$|{\rm RM}| > 100$~rad~m$^{-2}$. Thus the distribution of RMs which we have
measured argues for a distance to the emitting region in the 
range 1.3--4.5~kpc,
corresponding to pulsars with RMs which are similarly small and
negative.  Independent measurements further limit the distance to
the source of polarized emission: a lower limit of 2~kpc has been
inferred from \HI\ absorption towards polarized emission in this same
region of the sky (\cite{dic97}), while in 
Section~\ref{sec_disc_hii_dist} below we argue
from the depolarizing effect of \HII\ regions that this emission must
be produced between 3 and 6.5~kpc.  Taking into account all
these constraints, we find that all observations are consistent with
polarized emission in the Test Region originating at a mean distance
$3.5\pm1.0$~kpc.

Considering the spiral structure of the Galaxy in this direction
(e.g.\ \cite{gg76}; \cite{ch87b}), we find that the line-of-sight is
crossed by three spiral arms: the Carina arm at a
distance of $\sim$1.5~kpc, the Crux
arm at $\sim$3.5~kpc, and the Norma arm, which runs nearly parallel to
the line-of-sight in the range 6.5--10.5~kpc.  
Given that the diffuse component of synchrotron emission from
the Galaxy originates primarily in its spiral arms (\cite{bkb85}),
it thus seems reasonable to conclude from
our distance estimate that the polarization
we are detecting originates predominantly in the
Crux arm at 3.5~kpc. 

It is not surprising that we
do not detect polarization from the more distant Norma arm,
because this structure is largely
tangent to our line-of-sight. Internal Faraday rotation
within this arm will thus be significant, largely depolarizing
the emission from it (\cite{bur66}; \cite{sbs+98}). 
However, no such effects apply to the nearer Carina arm.  We can limit
the amount of polarization coming from this arm by considering
polarized emission in the direction of the \HII\ region RCW~94, which,
at a distance of 3~kpc (\cite{ch87b}), is behind the Carina arm. As
shown in Figs~\ref{fig_hii_pol} and \ref{fig_rcw94_slice}, and
discussed in Section~\ref{sec_disc_rcw} below, a halo surrounds RCW~94,
through which background emission is completely depolarized. The lack
of foreground polarization in this halo limits the relative
contribution of polarization being produced in the Carina arm to
$\la20$\% of that coming from the Crux arm. Because the Carina arm is
at less than half the distance of the Crux arm, comparable polarized
structures generated in it will have a correspondingly larger angular size,
and may thus go largely undetected by the interferometer.

\subsubsection{Canals of Low Polarization}

The polarization seen at 1.4~GHz is riddled with narrow canals
of reduced polarization, running randomly through it. Such
structures have been seen in various other 
wide-field images of radio polarization,
most notably in the data of Duncan \etal (1996\nocite{dshj96}),
Uyaniker \etal (1999\nocite{ufr+99}) and Haverkorn \etal
(2000\nocite{hkd00}). Haverkorn \etal (2000\nocite{hkd00})
specifically consider these canals, and show that in
their data, the polarization position angles on either sides of these
canals always differ by $90^\circ$. This demonstrates that
these canals are produced by beam depolarization, in
which perpendicular position angles are averaged together over
a resolution element to
result in no net polarization. Haverkorn \etal (2000\nocite{hkd00})
interpret these canals as being due to changes in the
foreground RM by just the amount to produce a $90^\circ$ rotation
in position angles.

It is beyond the scope of this paper to make a detailed
analysis of the properties of these
canals as was carried out by Haverkorn
\etal (2000\nocite{hkd00}).  However, it can be seen in
Fig~\ref{fig_ext_theta} that there is a sudden change in position
angle around the boundary of each patch of polarization --- most of these
sudden changes indeed correspond to canals. Furthermore, examination of
specific regions indicates
that most of the canals we see are one beamwidth wide, and
that at the positions of almost all of these canals, a sudden change in
position angle by $\sim90^{\circ}$ is seen. We demonstrate this in
Fig~\ref{fig_canals}, where we show the polarized intensities and
polarized position angles in a region near $(331\fdg5, +3\fdg0)$
containing several canals. Thus our observations support the
conclusion of Haverkorn \etal (2000\nocite{hkd00}), that canals occur
due to beam depolarization over adjacent regions of differing position
angle.

However, when one examines the change in RM across these canals, we
find that in general the RM does not noticeably change, and certainly
does not shift by the 33~rad~m$^{-2}$ required 
to rotate the position angle of polarization by 90$^\circ$
at 1.4~GHz.  This is apparent in
Fig~\ref{fig_canals}, where it can be seen that changes in RM across a
canal are no more significant than changes in RM seen within each
bright clump of polarized emission.  The canals we see are thus
possibly intrinsic to the source of polarized emission, and represent
emitting regions where the magnetic field geometry is significantly
non-uniform. These canals are not seen in the 2.4~GHz data shown in
Fig~\ref{fig_L_11cm}, presumably because of the much poorer angular resolution
of these data.

\subsection{Voids in Polarization}
\label{sec_disc_voids}

Figs~\ref{fig_ext_L} and \ref{fig_ext_L_2} demonstrate the
presence of two voids of polarization in the Test Region, each
a few square degrees in extent. These voids have the following
properties:

\begin{itemize}

\item They are approximately elliptical in shape.

\item They have no counterparts in total intensity.

\item The polarized intensity falls steadily to zero
as one considers lines of sight projected increasingly
close to their centers. This is demonstrated for void~1
in Fig~\ref{fig_void_slice}, where we plot the mean polarized
intensity in the range $330\fdg 0 < l < 332\fdg2$,
averaged between $b = 0\fdg7$ and $b=1\fdg2$.
This region is marked by a box in Fig~\ref{fig_ext_L_2}.
In the interior of these voids,
there is no detectable polarization down to a limit
$L \la 1.5$~mJy~beam$^{-1}$, a factor 3--4 less
than seen immediately outside the voids.

\item The voids are also seen at 2.4~GHz. At
this higher frequency,
the intensity of polarization
decreases by a factor of $\sim$2--3 in their
interiors relative to nearby
regions of bright polarization.

\item As one moves from the edges of the voids into their
interiors, the polarized intensity becomes progressively
fainter and clumpier.

\item Large rotation measures ($|{\rm RM}| > 150$~rad~m$^{-2}$)
are seen around the edges of these voids.

\end{itemize}

To the best of our knowledge, voids in polarization such as those described
here have not been previously reported.  Gray \etal
(1999\nocite{gld+99}) detected large regions of depolarization in a
polarimetric survey of the 2nd Galactic quadrant, carried out at the
same frequency and spatial resolution as for the SGPS. However the
voids which Gray \etal (1999\nocite{gld+99}) describe are coincident with,
and have a morphology which is strongly correlated with, the bright
\HII\ regions W3 and W4. The voids in the SGPS, on the other hand,
have no such counterparts detected in total intensity. Meanwhile,
Duncan \etal (1999\nocite{drrf99}) have recently presented
a survey of 2.7~GHz polarization in the 1st Galactic quadrant,
using the 100-m Effelsberg radiotelescope. They find
in their data that regions of low linear polarization coincide
with regions of enhanced \HI\ emission, and propose that these
\HI\ clouds are depolarizing emission propagating through them.
However, we have examined the SGPS \HI\ data for this region
(\cite{mgd+00}), and find no structures
or enhancements in \HI\ coincident with the voids.

We thus conclude that the voids seen in the SGPS
Test Region represent a previously unidentified
phenomenon. There are two possible explanations to account for them:
either they represent regions where the level of intrinsic polarization
is low, or they are the result of propagation through a foreground
object, whose properties have depolarized the emission at both 1.4 and
2.4~GHz.

We think the former possibility to be unlikely for two reasons.
Firstly, if the voids are intrinsic to the emitting regions, then the
distance of 3.5~kpc inferred in Section~\ref{sec_disc_dist} implies
that they are hundreds of parsecs across --- it is hard to see what
could produce such uniformly low polarized intensity across such large
regions. Secondly, for structure intrinsic to the emitting regions, we
expect that the morphology of polarization will either be correlated or
anti-correlated with structure seen in total intensity (the latter if
the voids are the result of internal depolarization), as discussed in
Section~\ref{sec_disc_origin} above. However, Fig~\ref{fig_I_11cm}
demonstrates that the appearance of the total intensity emission within
the voids is no different from elsewhere in the field.  We therefore
conclude that the voids cannot easily be explained as being intrinsic
to the emitting regions.

In further discussion, we consider in more detail the second possibility,
that the voids are caused by foreground depolarization.
There are several mechanisms through which such an
effect can occur
(see \cite{bur66}; \cite{tri91}; \cite{sbs+98}).
The first of these is bandwidth
depolarization, which occurs when there is
significant Faraday rotation across a single
spectral channel. This causes polarization vectors
at either edge of the channel to be out of phase,
and hence the polarized fraction to be reduced
when averaged across the channel. The
degree of bandwidth depolarization is given by
(e.g.\ \cite{gw66}):
\begin{equation}
p_b  = \left|\frac{\sin \Delta\Theta}{\Delta\Theta}\right|
\label{eqn_depol_band}
\end{equation}
where $p_b$ is the depolarization factor
as defined in Section~\ref{sec_disc_origin} above but due to bandwidth
effects, and
$\Delta \Theta = 2{\rm RM} c^2 \Delta\nu/\nu^3$ is the change in
angle across a spectral channel of width $\Delta\nu$.  An important
constraint on bandwidth depolarization comes from source 21 in
Table~\ref{tab_pol}, which is significantly linearly polarized despite
being deep within void 1.  The RM
measured for this source is $-762\pm12$~rad~m$^{-2}$, and it can be
seen in Fig~\ref{fig_rm_plot} that the fit to the data is very good
($\chi_r^2 = 1.2$). It is highly likely that the value measured is the
true RM for this source: if there were additional turns of Faraday
rotation between each spectral channel, this would require $|{\rm RM}|
> 5000$~rad~m$^{-2}$, which from Equation~(\ref{eqn_depol_band})
corresponds to a bandwidth-depolarization factor $p_b \approx 0.15$.
This would then imply that the intrinsic fractional polarization of
this source is $\sim45$\%, significantly greater than for any other
source in the Test Region or for extragalactic sources in general. We
thus regard it highly likely that there are no ambiguities in the RM
which we have measured.  

The measured RM of $-762$~rad~m$^{-2}$ then
differs by no more than $\sim$1000~rad~m$^{-2}$ from the RM value
predicted through the Galaxy at this position by Dickey \etal (in
preparation).  This difference is an upper limit on the
total Faraday depth of void 1 at this position.  We can thus
immediately rule out bandwidth depolarization as an explanation for the
voids, as a RM of $\pm$1000~rad~m$^{-2}$ produces negligible bandwidth
depolarization in our data ($p_b \sim 0.95$), and could
not produce the voids which we see.

A second cause of depolarization is a spatial gradient in RM.
Such a gradient will produce a variation in the 
angle of polarization as a function of position. If this
variation is significant when averaged across a single synthesized
beam, the polarized fraction can be greatly reduced.
Provided the RM gradient is resolved,
the amount of depolarization produced within a Gaussian synthesized
beam is given by (\cite{sbs+98}):
\begin{equation}
p_g = e^{-\frac{1}{\log_e 2}\left(\frac{d{\rm RM}}{dr}\right)^2 \lambda^4}
\label{eqn_depol_grad}
\end{equation}
where $\frac{d{\rm RM}}{dr}$ is the gradient of the RM across
the sky in units of rad~m$^{-2}$~beam$^{-1}$, and
$p_g$ is the depolarization factor due to gradient effects.
While only mild spatial gradients in RM,
of the order of $\sim25$~rad~m$^{-2}$~beam$^{-1}$, are required to 
cause significant depolarization at 1.4~GHz, for
any reasonable
geometry these gradients in RM will be most significant near
the edges of the structure (see discussion on RM gradients
in Section~\ref{sec_disc_rcw} below).
Fig~\ref{fig_void_slice} clearly
demonstrates that in fact depolarization is mildest near the
edges of the voids, and becomes severe in the interior. Thus it is
difficult to see how gradients in RM can account for the voids.

The remaining possibility is that depolarization in the voids is due to
beam depolarization, in which the RM varies randomly on small scales.
Depolarization occurs when linearly polarized emission
Faraday rotated with many different RMs (and hence different position
angles) is averaged over the beam.  Beam depolarization has the
specific property that it will depolarize extended emission, but does
not affect the polarized intensity of point sources. This can
thus account for the presence of the polarized source 21 within void 1.  
Furthermore, a characteristic property of mild beam depolarization is that
it induces a cell-like structure in the distribution of
polarized intensity (e.g.\ Moffett \& Reynolds 1994a,b\nocite{mr94,mr94b}),
just as is seen around the perimeters of the voids.

In the following discussion, we show that the
properties of void~1 can be understood using a simple
model of beam depolarization.
(We do not consider void~2 in detail because its perimeter
is less well-defined, but similar arguments apply.)

\subsubsection{A Model for Void 1}
\label{sec_disc_void_model}

We consider void 1 to be a caused by a 
sphere of uniform electron density $n_e$~cm$^{-3}$,
centered on $(332\fdg5, +1\fdg2)$ with a radius of
$1\fdg4$. If the distance to the sphere is $d$~kpc, this corresponds to a
spatial diameter $2R = 49d$~pc. The resolution of the observations is
$\sim$1~arcmin, corresponding to a spatial resolution
$\delta=0.3d$~pc.

Within the sphere, we suppose there are both random and ordered
components to the magnetic field.  The ordered component is uniform,
and has mean value $B_u$~$\mu$G, oriented at an angle $\theta$ to the
line of sight.
The random component has magnitude $B_r$~$\mu$G: we assume that the
random component is coherent within individual cells of size
$l$~pc, but that the orientation from cell to cell is random. Uniformly
polarized rays which propagate through a different series of cells will
experience differing levels of Faraday rotation, resulting in beam
depolarization when averaged over many different paths.

The level of depolarization resulting from this process
depends on two factors: the dispersion of RMs produced
by passage through the sphere, $\sigma_{\rm RM}$~rad~m$^{-2}$, and
the characteristic scale on which these fluctuations
occur, $\psi$~arcsec.

We first assume that the cells in the source are arranged
in a regular grid, so that two adjacent rays either
pass through the same sequence of cells, or through completely
independent sequences. In this case the scale for
RM fluctuations is equal to the cell size,
so that $\psi = 200l/d$~arcsec.
We can relate $\sigma_{\rm RM}$ to the properties of
an individual cell as follows. Consider radiation
passing through a single cell within the source.
The rotation measure produced by passage through this
cell is given by:
\begin{equation}
{\rm RM}_i = K n_e l (B_r \cos \phi_i + B_u \cos \theta),
\end{equation}
where $\phi_i$ is the angle to the line of sight of the
random component of the field within that cell.
The total RM after passage through N cells is then:
\begin{equation}
{\rm RM} = \sum_{i=1}^{N} {\rm RM_i} = K n_e l \left( N B_u \cos\theta +
B_r \sum_{i=1}^{N} \cos \phi_i \right)~{\rm rad~m}^{-2}.
\end{equation}
The $N$ values of $\cos \phi_i$ are independently  distributed with
a mean of 0 and a variance of $1/3$. Then using the
central limit theorem, we find that for large $N$, $\sum_{i=1}^{N} \cos
\phi_i$ is normally distributed with mean 0 and variance $N/3$.
Thus we expect:
\begin{equation}
<{\rm RM}>~=~K n_e D  B_u \cos\theta~{\rm rad~m}^{-2}
\label{eqn_mean_rm}
\end{equation}
and
\begin{equation}
\sigma_{\rm RM} = K n_e \frac{B_r}{\sqrt{3}} \sqrt{Dl}~{\rm rad~m}^{-2},
\end{equation}
where $D = Nl$~pc is the depth through the source at a particular
position. 

The above result assumes that adjacent patches of different
RM correspond to completely independent paths through the source.
However, it is more realistic to assume that the cells do not
fall in ordered columns, but that each layer of cells is
randomly aligned with respect to every other layer. In this case,
the values of RM in adjacent patches are correlated, since two
neighboring paths will have some cells in common, and we thus
expect a smaller value of $\sigma_{\rm RM}$ than in the
case of an ordered grid of cells. Furthermore, we expect
the scale of RM fluctuations to be smaller, since one no
longer needs to move a full cell across the source before
encountering a region of differing RM.

We have carried out Monte Carlo simulations to quantify
how $\psi$ and $\sigma_{\rm RM}$ change in the case
of randomly aligned cells. We find that the characteristic
scale of RM fluctuations is decreased by a factor of two,
corresponding to an angular scale half the cell size, so that
$\psi = 100 l /d~{\rm arcsec}$.
Furthermore, we find that the RMs are still normally distributed,
but with a dispersion smaller by a factor of 2, so that
\begin{equation}
\sigma_{\rm RM} = K n_e \frac{B_r}{2\sqrt{3}} \sqrt{Dl}~{\rm rad~m}^{-2}.
\label{eqn_sigma_rm}
\end{equation}

For particular values of $\psi$ and $\sigma_{\rm RM}$,
Tribble (1991\nocite{tri91}) has shown that the resulting
depolarization factor due to beam depolarization is:
\begin{equation}
p_f^2 \approx \frac{1-e^{\left(-\kappa^2-4\sigma_{\rm RM}^2\lambda^4\right)}}
{1 + 4\sigma_{\rm RM}^2\lambda^4/\kappa^2}
+ e^{\left(-\kappa^2 - 4\sigma_{\rm RM}^2\lambda^4 \right)}
\label{eqn_beam_depol}
\end{equation}
where $\kappa = 5.72\times10^{-3} \psi d / \delta$.
For increasingly smaller scales for the RM
fluctuations (i.e.\ smaller values of $\kappa$ and $\psi$),
depolarization becomes increasingly severe (i.e.\
$p_f$ decreases).
In the limiting case that 
$\kappa \rightarrow 0$, this
expression reduces to (\cite{bur66}; \cite{tri91}):
\begin{equation}
p_f = e^{-2\sigma_{\rm RM}^2\lambda^4}.
\label{eqn_beam_depol_2}
\end{equation}

For a spherical geometry we have that $D = 2\sqrt{R^2-r^2}$~pc, 
where $r$~pc is the impact parameter of the line of sight relative
to the center of void 1.
We can then adopt a particular value of $\psi$ and
then use Equations~(\ref{eqn_sigma_rm}) and (\ref{eqn_beam_depol}) 
to fit to the profile of polarized intensity seen in Fig~\ref{fig_void_slice}.

However an important constraint on any model is that significant
beam depolarization is also seen within void 1 at 2.4~GHz.
At the much poorer resolution of the 2.4~GHz data (10.4 arcmin), we can
use Equation~(\ref{eqn_beam_depol_2}) to predict
the degree of depolarization seen at 2.4~GHz.
The depolarization seen at 2.4~GHz within void~1 has value $p_f \sim 0.3-0.5$,
implying peak dispersions in the
RM fluctuations of $\sigma_{\rm RM} \sim 35-50$~rad~m$^{-2}$.

Given this constraint, we find that the only parameters which can reasonably
fit the 1.4~GHz polarization profile seen in Fig~\ref{fig_void_slice}, while
simultaneously  producing the appropriate levels of depolarization at
2.4~GHz, are $\psi \approx 60'' - 80''$ (i.e.\ $l \approx [0.6-0.7]d$~pc
and $\kappa \approx 1.1-1.5$) and $n_e B_r \approx 30 / d$~$\mu$G~cm$^{-3}$.
The 1.4~GHz polarized intensity predicted by this model is plotted
in Fig~\ref{fig_void_slice}.
Shifting the assumed center and radius of the sphere by
reasonable amounts changes these parameters by less than 10\%.
We know that $d < 3.5$~kpc (since this is the distance to the
polarized emission behind the void). Thus
a joint lower bound on the density and random component
of the field within the sphere
is $n_e B_r \ga 9$~$\mu$G~cm$^{-3}$.

We can limit the electron density in the sphere by noting
that the 2.4~GHz surface brightness in this direction
in the survey of Duncan \etal (1995\nocite{dshj95})
is $\sim4$~K. This is then an upper limit on the
brightness temperature of any free-free emission produced
by the sphere. For optically thin emission,
and assuming a typical electron temperature
$T_e = 8000$~K, an upper limit
on the emission measure through the center 
of the sphere (e.g.\ \cite{mh67})
is then EM~$=\int n_e^2 dl < 7000$~pc~cm$^{-6}$. This
limit implies that $n_e < 12d^{-0.5}$~cm$^{-3}$,
which together with the fit to the polarization
profile described above implies
$B_r > 2.5d^{-0.5}$~$\mu$G~$ \ga 1.3$~$\mu$G.

We can constrain the strength of the ordered component of the field by
noting that near the edges of the voids, where gradients in RM are most
severe, depolarization due to such gradients is negligible.
Specifically, the behavior of polarized intensity in
Fig~\ref{fig_void_slice} is such that we can approximate $p_g > 0.5$ at
$r = R - \delta$.  Using the fact that the gradient
in RM is given by:
\begin{equation}
\frac{d{\rm RM}}{dr} = \frac{-2 K n_e B_u r \delta \cos \theta}
{\sqrt{R^2 - r^2}},
\label{eqn_rm_gradient}
\end{equation}
we can use Equations~(\ref{eqn_depol_grad})
and (\ref{eqn_mean_rm}) to determine that
$n_e B_u \cos \theta \la 4.9/d$~$\mu$G~cm$^{-3}$. The peak RM
through the void is then $\la$200~rad~m$^{-2}$, consistent with
the limit $|{\rm RM}| < 1000$~rad~m$^{-2}$ inferred from the total RM
of source 21 as discussed in Section~\ref{sec_disc_voids} above.

If we assume that the strengths of the uniform
and the random components of the field are approximately equal
(e.g.\ \cite{jon89}; \cite{gh94c}), then our combined limits on $B_u$ and $B_r$
imply that $\cos \theta \la 0.15$, so that $B_u$ is largely in
the plane of the sky.

\subsubsection{Interpretation of the Model}
\label{sec_disc_void_inter}

While the above model is simplistic in its assumptions, it successfully
reproduces the polarization profile at 1.4~GHz, accounts for
depolarization seen also at 2.4~GHz, and explains the fact that source
21 is not depolarized nor has an excessive RM. 

While voids in polarization such as described here have not been
previously reported, localized regions of similarly enhanced RM
fluctuations have been seen in polarization studies of extragalactic
sources. In particular, Simonetti \& Cordes (1986\nocite{sc86}) saw
fluctuations in RM on arcmin scales at the level of $\sigma_{\rm RM}
\la 5$~rad~m$^{-2}$ over most of the sky, but found enhanced
fluctuations of magnitude $\sigma_{\rm RM} \approx 40$~rad~m$^{-2}$ in
a region of the Galactic Plane near $l=90^{\circ}$.  Similarly, Lazio,
Spangler \& Cordes (1990\nocite{lsc90}) measured the RMs of sources
behind the Cygnus OB1 association, and found RM fluctuations on arcmin
scales at the level of $\sigma_{\rm RM} \sim 25$~rad~m$^{-2}$.

It has been suggested that the high dispersion in RM in such regions is
the result of either an expanding wind bubble or supernova blast wave,
either of which will sweep a dense shell of material from its
environment.  Propagation through the turbulence and instabilities
produced in this surrounding shell will then generate the observed
fluctuations in RM (\cite{sc86}; \cite{sc98}). However, because the
severity of beam depolarization increases with the line-of-sight pathlength
through a source, the polarization profile seen in
Fig~\ref{fig_void_slice} cannot be easily produced by a thin shell, for
which we would expect to see significantly more beam depolarization
around the perimeter than through the center.  Furthermore, one might
expect to see the swept up material in \HI, and yet examination of the
SGPS \HI\ data for this region shows no such features in this region.

We therefore argue that the voids we observe are best explained as
structures which are turbulent throughout their extent, rather than in a
shell around their perimeters. We propose that the voids are caused by low
density \HII\ regions, which have extents, densities and magnetic
fields similar to those inferred here (e.g.\ \cite{hc80}; \cite{lph96};
\cite{val97}). This interpretation is supported by the presence of
significant levels of diffuse H$\alpha$ emission coincident with void~2
(\cite{abg+91}; \cite{gag+94}). 
There are multiple overlapping H$\alpha$ complexes in
this low-latitude region, so it is not possible to uniquely identify an
\HII\ region which might be specifically associated with void~2.
However, we note that most of the H$\alpha$ nebulae in this region are
at distances of between 1 and 3~kpc (\cite{gag+94}), placing them in
front of the polarized background as required in our model.

Unfortunately, no information on H$\alpha$ is currently available for
void~1. We can estimate an approximate distance to this region by
noting that turbulence in \HII\ regions is  consistently seen to
have an outer scale of $\sim0.15$~pc (\cite{mjd95}; \cite{jon99}).
Requiring this to be comparable to the scales inferred for turbulent
cells in void 1, we can infer an approximate distance to void 1 of
$d\sim0.25$~kpc.  

It is interesting to note that the O9V star
HD~144695, at $(332\fdg25, +1\fdg30)$, is very close to the projected
center of void 1, and is at a distance of $0.30\pm0.16$~kpc
(\cite{gcc82}; \cite{plk+97}).  If we suppose that void~1 is a
Str\"{o}mgren sphere associated with this star, its radius of
$7.3\pm3.9$~pc implies a density $n_e \sim 25\pm20$~cm$^{-3}$
(\cite{pth69}), which is consistent with the limit $n_e < 12d^{-0.5}
\sim 20$~cm$^{-3}$ inferred above from the upper limit on emission
measure.

Two properties of the voids which our simple model cannot account for
are the requirement that the uniform component of the magnetic field be
largely oriented in the plane of the sky, but that we generally observe
coherent regions of large RM (of the order of a few hundred
rad~m$^{-2}$) around the edges of the voids. We suggest that both these
results can be explained if the uniform component of the magnetic field
within the void is oriented so that its radial component (relative
to the center of the void) is always
zero.  This will result in a magnetic field perpendicular to the line
of sight over most of the void, but which is parallel to the line of
sight (and can thus potentially produce high RMs) around the perimeter.
If the ambient magnetic field is well-ordered, such a field
geometry will naturally arise during the expansion phase of an
\HII\ region as it interacts with surrounding material. The
amount of material swept up by this process would be much smaller
than in the case of a rapidly expanding wind bubble or SNR, 
consistent with the lack of detection of an \HI\ shell surrounding
the void.

To summarize, the properties of void 1 can be explained if it is an
\HII\ region of density $n_e \sim 20$~cm$^{-3}$ and magnetic field $B_r
\approx B_u \sim 5$~$\mu$G at a distance $d \sim 0.3$~pc,
potentially associated with the O9V star HD~144695.
The depolarization which we observe is caused by turbulence on
scales $\sim0.2$~pc, with dispersion in RM $\sigma_{\rm RM} =
35-50$~rad~m$^{-2}$.  

Considerable further study will be required before we are
able to properly ascertain the nature of the voids. In the
particular case of void 1, H$\alpha$ observations are
highly desirable, both to better constrain the emission
measure through the source and to determine a kinematic
distance for it from the velocity of the line emission.
More generally, it is likely that the full SGPS
will allow us to identify a large number of such voids,
which will allow us to better determine
the properties of the population.

\subsection{\HII\ regions}
\label{sec_disc_hii}

\subsubsection{Depolarization seen towards RCW~94}
\label{sec_disc_rcw}

As seen in Figs~\ref{fig_hii_pol} and \ref{fig_rcw94_slice}, 
the \HII\ region RCW~94
shows reduced linear polarization in its interior, and is
furthermore surrounded by a halo in which no linear polarization
is detected.  We take these results to
indicate that this source is depolarizing background emission
propagating through it, as was seen towards the \HII\ complex W3/W4
by Gray \etal (1999\nocite{gld+99}).  We now attempt to account
quantitatively for this behavior.

The region of depolarization seen towards RCW~94 extends beyond the
boundaries of the source as seen in total intensity (indeed this is
where depolarization is most extreme).\footnote{This is also beyond the
extent of the source as seen in H$\alpha$, for which the size and shape
are similar to those seen in the radio (\protect\cite{gag+94}).} This
indicates that the full extent of the \HII\ region is larger than that
part of it which produces detectable emission. In subsequent
discussion, we assume the extent of the source to be a circle of radius
16~arcmin, corresponding to the extent of the depolarized region.

The various mechanisms by which a foreground source can depolarize
emission propagating through it were discussed in
Section~\ref{sec_disc_voids}.  Since \HII\ regions are not expected to
be sources of polarization themselves, we can immediately rule out
internal depolarization.  In the following discussion, we assume some
simple geometries for RCW~94, and see if the polarized intensity
observed towards it can then be understood in terms of bandwidth, gradient
and beam depolarization.  The levels of depolarization which we are
trying to account for are $p \sim 0.5$ in regions coincident with
bright radio continuum, and $p \la 0.2$ in a halo surrounding this
region.

We first assume for RCW~94 the same geometry as for void~1 discussed
above:  a sphere of radius $R$~pc at distance $d$~kpc observed at a
spatial resolution $\delta$~pc, within which the electron density has a
uniform value of $n_e$~cm$^{-3}$.  In this particular case, we have
that $d = 3$~kpc, $2R = 28$~pc, $\delta = 0.9$~pc, and $n_e =
20$~cm$^{-3}$ (\cite{smn83}; \cite{ch87b}).  As in 
Section~\ref{sec_disc_void_model} above,
we assume a magnetic field with a
uniform component $B_u$, inclined at an angle $\theta$ to the line of
sight, and with a random component $B_r$, characterized by a turbulent
scale $l$.

To determine the total depolarization, we combine
Equations (\ref{eqn_depol_band}), 
(\ref{eqn_depol_grad}) and (\ref{eqn_beam_depol}), so that:
\begin{equation}
p = p_b~p_g~p_f.
\label{eqn_depol_tot}
\end{equation}
Because the 2.4~GHz data are of insufficient spatial resolution
to provide useful information on this region, we cannot
uniquely constrain the parameters of the source
as was done for void 1 in 
Section~\ref{sec_disc_void_model}. However,
a general feature of any set of parameters is that the terms
in $p_b$ and $p_g$ increase (i.e.\ cause less depolarization)
with increasing $r$, while $p_g$ decreases with increasing $r$.
An example of this behavior is shown in the upper
panel of Fig~\ref{fig_rcw_sphere}, where we
show the expected polarization profile for reasonable
parameters $B_r = B_u = 1.2$~$\mu$G,
$l = 0.15$~pc and $\theta = 60^\circ$. In this case
the peak RM through the source is 270~rad~m$^{-2}$, so that
bandwidth depolarization is negligible. Beam depolarization
is most severe in the center,
but becomes increasingly less significant at larger radii. 
Depolarization due to RM gradients is only significant
in the outer half of the source,
and completely depolarizes the emission
beyond $r/R \ga 0.7$. The net polarized intensity then
qualitatively agrees with observations,
in that it produces $p \sim 0.5$ near the center of
the source and $p \la 0.2$ around its edges.

However, this model is at odds with the shape of the
observed polarization profile seen in Fig~\ref{fig_rcw94_slice},
which has approximately constant surface brightness
at small radii, and suddenly decreases
around the outer edges of the source. We can find
no set of parameters for a spherical geometry which
can reproduce this behavior.
Note that if we relax the assumption that $n_e$ and $B_u$ are
uniform, and rather model either parameter as decreasing with
radius, then the gradients in RM are more severe, and $p_f$
will fall to zero for values of $r/R$ smaller than observed.

Thus in order to reproduce the observed behavior, we must
abandon a spherical model, and instead
consider a geometry in which the RM is approximately
constant as a function of radius for $r/R < 0.7$
(so that $p_b$ and $p_f$ are constant and
$p_g$ is unity), but which then decreases linearly to zero
for $0.7 < r/R < 1$ (so that $p_g$ suddenly decreases
to zero around the edges). 
A reasonable such geometry is a cylinder viewed
face-on, in the interior of which $n_e$ is constant, but around the
edges of which $n_e$ falls linearly to zero. 
Such a model produces a
constant emission measure (EM) for $r/R<0.7$, in agreement with the
approximately constant surface brightness in H$\alpha$ seen in this
region (\cite{gag+94}). The model also results in a low EM
around the outer edges of the source, which can account for the fact
that the size of the source as seen in depolarization is larger than
that delineated by its radio continuum and H$\alpha$ emission.

When we model this geometry
using the same parameters as in the spherical case above ($n_e =
20$~cm$^{-3}$, $B_u = B_r = 1.2$~$\mu$G, $l = 0.15$~pc and $\theta =
60^\circ$), and assuming that the depth through the source is equal to
its diameter projected onto the sky, we obtain a predicted polarization
profile as shown in the lower panel
of Fig~\ref{fig_rcw_sphere}.  This now correctly
reproduces the observed polarization properties of the source.

\subsubsection{Interpretation of the Model}

The fall-off in $n_e$ which we have invoked to account for the depolarized
halo around RCW~94 is suggestive of similar features seen
in polarimetric observations of the \HII\ regions W3, W4 and W5
(\cite{gld+99}). Not only do these
sources depolarize the emission propagating through them in a
similar way as seen through the interior of RCW~94, but there are also
suggestions that the depolarization extends beyond the
boundaries of these sources as defined by their continuum radio
emission.  Gray \etal (1999\nocite{gld+99}) interpret
this to indicate that these \HII\ regions are embedded in
low density cocoons, the existence of which has
been previously inferred from recombination line
emission seen both in the vicinity of specific \HII\ regions and also
along the entire Galactic Plane (\cite{hp76};
\cite{ana85}; \cite{acc87}).

However, the presence of significant CO emission at the same position
and systemic velocity as for RCW~94 (\cite{bact89}) suggests that the
\HII\ region may rather be interacting with a molecular cloud.  This
possibility is supported by \HI\ observations of the region, which show
that RCW~94 is embedded in a shell of \HI\ emission, which is further
surrounded by a ring of decreased \HI\ emission (\cite{mdg+00b}).
McClure-Griffiths \etal\ (2000d\nocite{mgd+00}) argue that this structure in
\HI\ confirms that RCW~94 is embedded in a molecular cloud, the shell
of emission resulting from H$_2$ molecules dissociated by the
\HII\ region, and the surrounding region of reduced
\HI\ corresponding to regions of undisturbed molecular material.
Simulations of \HII\ regions evolving within molecular clouds
(\cite{rtf95} and references therein) show that for certain forms of
the density profile within the parent cloud, the shock driven into the
cloud by the embedded expanding \HII\ region can produce a halo of
partially ionized material around the latter's perimeter, which
would produce the fall-off in $n_e$ required to produce the
depolarization observed in Figs~\ref{fig_hii_pol}
and~\ref{fig_rcw94_slice}.

We note that high RMs are seen around the perimeter
of RCW~94 in Fig~\ref{fig_ext_rm}. This may be due
to the ambient magnetic field ``wrapping around''
the expanding \HII\ region as was proposed 
in Section~\ref{sec_disc_void_inter} for the voids.
Alternatively, these high RMs
may simply be due to the magnetic fields and ionized
material in which the entire RCW~94 complex 
(see e.g.\ \cite{gag+94}) is embedded.

\subsubsection{Limits on the distance to the polarized emission}
\label{sec_disc_hii_dist}

The fact that RCW~94 (and possibly also G326.7+0.5) show this depolarizing
effect implies that the diffuse background polarization must
be propagating through these sources, and hence must be behind
them. This puts a lower limit of $\sim$3~kpc on the distance
to the source of the diffuse polarized emission in this direction.

Of the other $\sim$30 \HII\ regions in the field, most are either in
large voids of depolarization, or have sufficiently low densities
and diameters that they are unlikely to produce
significant depolarization.
The one exception this this is the \HII\ region G331.6--0.0, which
seems to have no effect on polarization at its position,
but which should produce significant
depolarization. 

G331.6--0.0 has $n_e = 250$~cm$^{-3}$ (\cite{smn83}) and
is at a distance $d=6.5$~kpc (\cite{cmr+75}; \cite{ch87b}),
corresponding to a
linear extent $2R = 30$~pc.  Given that magnetic field and
density are generally correlated
(\cite{th86}; \cite{val97}), it is unlikely that the magnetic
field is any less than $B \sim 1$~$\mu$G as assumed for the
lower density source RCW~94.  We therefore
expect significant ($p < 0.1$) depolarization through the
center of the source, due to both bandwidth and beam effects. In fact,
we observe in this region that $p \sim 1$.  The simplest interpretation of
this discrepancy is that this source is behind the polarized emission
seen at its position.  This limits the distance of the polarization
background to be closer than 6.5~kpc.

\section{Conclusions}

The ATCA's sensitivity, spatial resolution and spectral flexibility
have allowed us to study linear polarization and Faraday rotation in
the SGPS Test Region in unprecedented detail.
Even though this field covers less than 7\% of the full survey, a
variety of structures are apparent from this preliminary study. 

We identify the following main features seen in polarization:

\begin{itemize}

\item Diffuse linear polarization seen throughout the
Test Region, with structure on scales of degrees, and
completely uncorrelated with emission seen in total intensity.
The brightest polarized features
most likely represent intrinsic structure in the source
of emission, but fainter components are best
explained as being imposed by 
Faraday rotation of a uniformly polarized
background by foreground material.
The rotation measures determined towards this emission
generally lie between --100 and 0 rad~m$^{-2}$, which
indicates that this emission originates at a distance
of $\sim$3.5~kpc, most likely in the Crux spiral
arm of the Galaxy.

\item Two large regions devoid of polarization, each several degrees in
extent, and with no counterparts in total intensity. We have considered
one of these voids in detail, and have shown that the depolarization
observed through it can be explained by turbulence within it
on scales $\sim0.2$~pc.  We find that the properties of the void are
consistent with it being a nearby ($\sim250$~pc) \HII\ region of
density $n_e \sim 20$~cm$^{-3}$ and magnetic field $B \sim 5$~$\mu$G,
possibly associated with the O star HD 144695.

\item A halo of depolarization surrounding the \HII\
region RCW~94. We interpret this as delineating a
region of lower electron density surrounding
the \HII\ region. We propose that this region
corresponds to a halo of ionized material
resulting from the interaction of RCW~94
with a surrounding molecular cloud.

\end{itemize}

The results presented here demonstrate the value of interferometric
polarimetry as a probe of the ISM.  The full SGPS will allow us to
accumulate many more examples of the types of features we have described
here, with which we will be able to better determine their properties
and distributions.  Furthermore, we expect that the large-scale
properties of polarized emission will be a function of position within
the Galactic Plane (e.g.\ Duncan \etal 1997a,
1999\nocite{dhjs97,drrf99}), reflecting differing orientations of the
spiral arms with respect to the line of sight. Overall, we anticipate
that the SGPS will provide the opportunity for a comprehensive
study of the magneto-ionic medium of the inner Galaxy, on scales
ranging from sub-parsec turbulence up to the global structure of the
spiral arms and disc.

\begin{acknowledgements}

We thank Veta Avedisova for supplying us with her extensive catalogue
of star formation regions, and are grateful to Alyssa Goodman, Carl
Heiles, Larry Rudnick and Ellen Zweibel for stimulating discussions on
Faraday rotation, depolarization and turbulence. This research has made
use of the NASA Astrophysics Data System and the CDS SIMBAD database.
The Australia Telescope is funded by the Commonwealth of Australia for
operation as a National Facility managed by CSIRO.  This research has
made use of NASA's Astrophysics Data System Abstract Service and of the
SIMBAD database, operated at CDS, Strasbourg, France.  B.M.G.
acknowledges the support of NASA through Hubble Fellowship grant
HST-HF-01107.01-A awarded by the Space Telescope Science Institute,
which is operated by the Association of Universities for Research in
Astronomy, Inc., for NASA under contract NAS 5--26555. N.M.Mc-G. and
J.M.D.  acknowledge the support of NSF grant AST-9732695 to the
University of Minnesota. N.M.Mc-G. is supported by NASA Graduate
Student Researchers Program (GSRP) Fellowship NGT 5-50250.

\end{acknowledgements}


\bibliographystyle{apj1}
\bibliography{journals,modrefs,psrrefs,crossrefs}

\clearpage

\clearpage

\begin{table}[hbt]
\caption{Compact sources showing linear polarization.}

\begin{tabular}{ccccccc} \hline \hline
Source No. &   l   &  b   &   $L$ (mJy) & $I$ (mJy) & $L/I$ (\%)  &  RM (rad~m$^{-2}$) \\        
  (1)  &  (2) &   (3)       &  (4)      &  (5)        &   (6) &  (7) \\ \hline
1 & 325.81 &  +1.08 &  9.7 & 616.6 &  1.6 & $-23\pm11$ \\
2 & 325.93 &  +2.92 &  4.6 & 51.1  &  9.0 & $-159\pm28$ \\
3 & 325.96 &  +3.14 &  6.2 & 357.1 &  1.7 & $-172\pm20$ \\
4 & 326.30 &  +2.17 & 10.4 & 234.0 &  4.4 & $-357\pm10$ \\
5 & 326.54 &  +2.18 &  7.2 & 23.5  & 30.6 & $-505\pm14$ \\
6 & 326.80 &  +1.57 & 14.3 & 226.4 &  6.3 & $-768\pm8$ \\
7 & 326.94 &  +2.40 &  3.5 & 128.8 &  2.7 & $-808\pm29$ \\
8 & 327.31 &  +0.88 &  6.8 & 288.5 &  2.4 & $-156\pm16$ \\
9 & 327.52 &  +2.70 &  6.8 &  64.9 & 10.5 & $-595\pm16$ \\
10 & 327.82 &  +1.29 &  7.3 &  92.6 &  7.9 & $-882\pm14$ \\
11 & 327.87 &  +1.60 &  5.4 & 122.0 &  4.4 & $-818\pm21$ \\
12 & 327.99 &  +1.48 &  3.3 & 154.0 &  2.1 & $-1344\pm27$ \\
13 & 328.60 &  +3.05 &  6.0 & 136.3 &  4.4 & $-326\pm18$ \\
14 & 328.62 &  +1.52 &  3.8 &  63.6 &  6.0 & $-933\pm30$ \\
15 & 328.80 &  +2.80 &  4.6 &  85.9 &  5.4 & $-350\pm25$ \\
16 & 329.20 &  +2.83 &  9.9 &  57.3 & 17.4 & $-348\pm11$ \\
17 & 330.35 &  +1.81 &  3.3 &  89.9 &  3.7 &  $+25\pm35$ \\
18 & 330.56 &  +1.74 & 10.3 & 159.4 &  6.5 &   $+4\pm11$ \\
19 & 330.69 &  +1.63 &  3.9 &  12.3 & 32.0 & $+145\pm25$ \\
20 & 330.77 &  +2.92 & 22.1 & 297.3 &  7.4 & $+172\pm5$ \\
21 & 332.14 &  +1.03 &  9.4 & 140.0 &  7.1 & $-762\pm12$ \\ \hline \hline
\end{tabular}
\label{tab_pol}
\end{table}

\vspace{1cm}

\clearpage

\begin{figure}[htb]
\vspace{5cm}
\caption{Total intensity, $I$, in the SGPS Test Region.
The greyscale runs from --30 to +120 mJy~beam$^{-1}$. The
RMS in regions devoid of emission is 1.0--1.5~mJy~beam$^{-1}$.}
\label{fig_ext_I}
\end{figure}

\begin{figure}[htb]
\vspace{5cm}
\caption{Total intensity, $I$, in the SGPS Test Region as seen by the
2.4~GHz single dish survey of Duncan \etal
(1995\protect\nocite{dshj95}). The resolution is $10\farcm4$ and the
greyscale runs from --0.2 to +17~Jy~beam$^{-1}$, using a square-root
transfer function to better show faint structure.  The data shown here
correspond to the ``small-scale'' image of Duncan \etal
(1995\protect\nocite{dshj95}), in which the largest scale structure has
been filtered out.}
\label{fig_I_11cm}
\end{figure}

\begin{figure}[htb]
\vspace{5cm}
\caption{Linearly polarized intensity, $L = (Q^2 + U^2)^{1/2}$.
The greyscale runs from 0.4 to 9.5~mJy~beam$^{-1}$.}
\label{fig_ext_L}
\end{figure}

\begin{figure}[hbt]
\caption{As in Fig~\ref{fig_ext_L}, but
with a greyscale range 1.0 to 5.0~mJy~beam$^{-1}$.
The labeled ellipses delineate the approximate
extents of voids 1 and 2, while the box shows the
region of the field over which the profile
shown in Fig~\ref{fig_void_slice} was generated.}
\label{fig_ext_L_2}
\end{figure}

\begin{figure}[hbt]
\caption{Position angle (averaged across the band)
of linearly polarized intensity
$\Theta = \frac{1}{2} \tan^{-1}(U/Q)$. The values of the position
angles have been converted from equatorial to Galactic coordinates,
so that $\Theta = 0^\circ$ represents a vector pointing in
the direction of increasing $b$ (i.e.\ up),
and $\Theta$ increases in the direction of increasing $l$ (i.e.\
counter-clockwise). The greyscale
runs from $-40^\circ$ to $+115^\circ$.}
\label{fig_ext_theta}
\end{figure}

\begin{figure}[hbt]
\caption{Linearly polarized intensity, $L$, 
as seen by the 2.4~GHz single dish survey of Duncan \etal
(1997a\protect\nocite{dhjs97}), covering the same part
of the sky as in the SGPS Test Region.
The resolution is $10\farcm4$ and the greyscale runs 
from +1 to +360~mJy~beam$^{-1}$.}
\label{fig_L_11cm}
\end{figure}

\begin{figure}[htb]
\caption{Distribution of rotation measure across the SGPS Test
Region. The filled and open boxes represent positive and
negative RMs respectively. For $|{\rm RM}| < 150$~rad~m$^{-2}$,
the length of the side of each box is proportional to the RM at that position;
for  RMs of larger magnitudes, the boxes drawn
correspond to $|{\rm RM}| = 150$~rad~m$^{-2}$. At positions where
no box is shown, an RM could not be calculated
either due to insufficient signal-to-noise or a poor fit
to the position angles; RMs were also blanked at positions
corresponding to polarized point sources. Each RM shown corresponds
to the average over a $15\times15$ pixel region (i.e.\ $5'\times5'$);
only every second such RM along each axis is plotted.
The background image represents the linearly polarized intensity
from the Test Region as in Figs~\ref{fig_ext_L} and \ref{fig_ext_L_2},
but with a greyscale range of 0.2 to 19.0~mJy~beam$^{-1}$.}
\label{fig_ext_rm}
\end{figure}

\clearpage

\begin{figure}[htb]
\vspace{2cm}
\centerline{\psfig{file=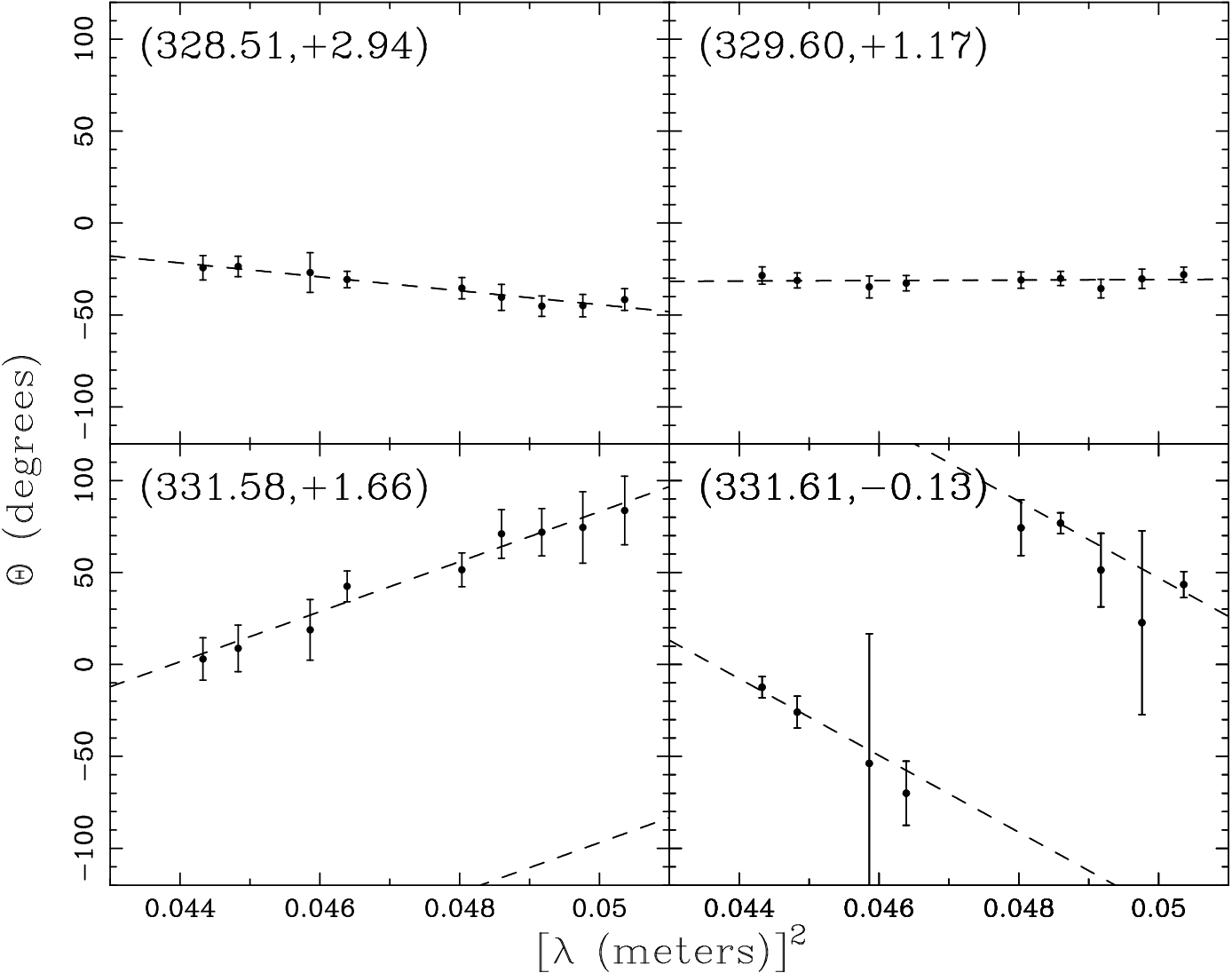,width=18cm,angle=270}}
\caption{Plots of polarization position angle, $\Theta$, 
against $\lambda^2$ towards four
regions of diffuse polarization; the coordinates
at which each measurement was taken are indicated in the
upper left corner of each panel.
The dashed line represents
the least-squares fit to the data of the function
$\Theta = \Theta_0 + {\rm RM}~\lambda^2 + n\pi$,
where $n=0, \pm1, \pm2, \ldots$;
the corresponding values of RM are (ordered
by increasing Galactic longitude) 
$-66\pm17$, $+2\pm13$, $+237\pm39$ and $-364\pm23$~rad~m$^{-2}$.
Note that the position angles shown here are relative to
equatorial, rather than Galactic coordinates; this has
no effect on the resultant RMs.}
\label{fig_rm_plot_2}
\end{figure}

\begin{figure}[htb]
\centerline{\psfig{file=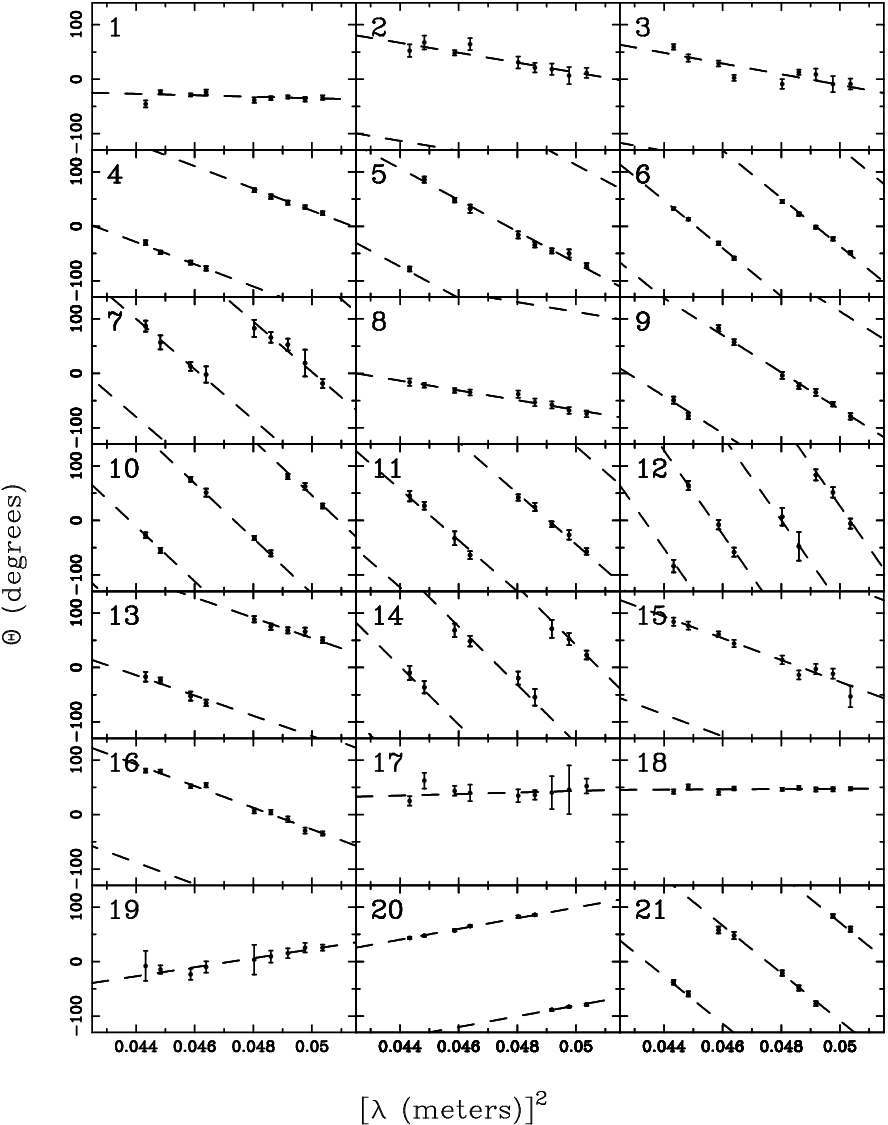,width=16cm,angle=270}}
\caption{Plots of polarization position angle against
$\lambda^2$ for
the 21 compact sources showing linear polarization as listed
in Table~\ref{tab_pol}. Details of the plots are as
in Fig~\ref{fig_rm_plot_2}.}
\label{fig_rm_plot}
\end{figure}

\clearpage

\begin{figure}[htb]
\vspace{2cm}
\caption{Polarized emission in the vicinity of the \HII\
regions RCW~94 (upper right) and G326.7+0.5 (lower left).
The greyscale represents linearly
polarized intensity,
while the single contour corresponds to 
total intensity emission from the same region at
the level of 45~mJy~beam$^{-1}$.
The near-horizontal line 
corresponds to the intensity slice shown in Fig~\ref{fig_rcw94_slice}.}
\label{fig_hii_pol}
\end{figure}

\begin{figure}[htb]
\vspace{2cm}
\centerline{\psfig{file=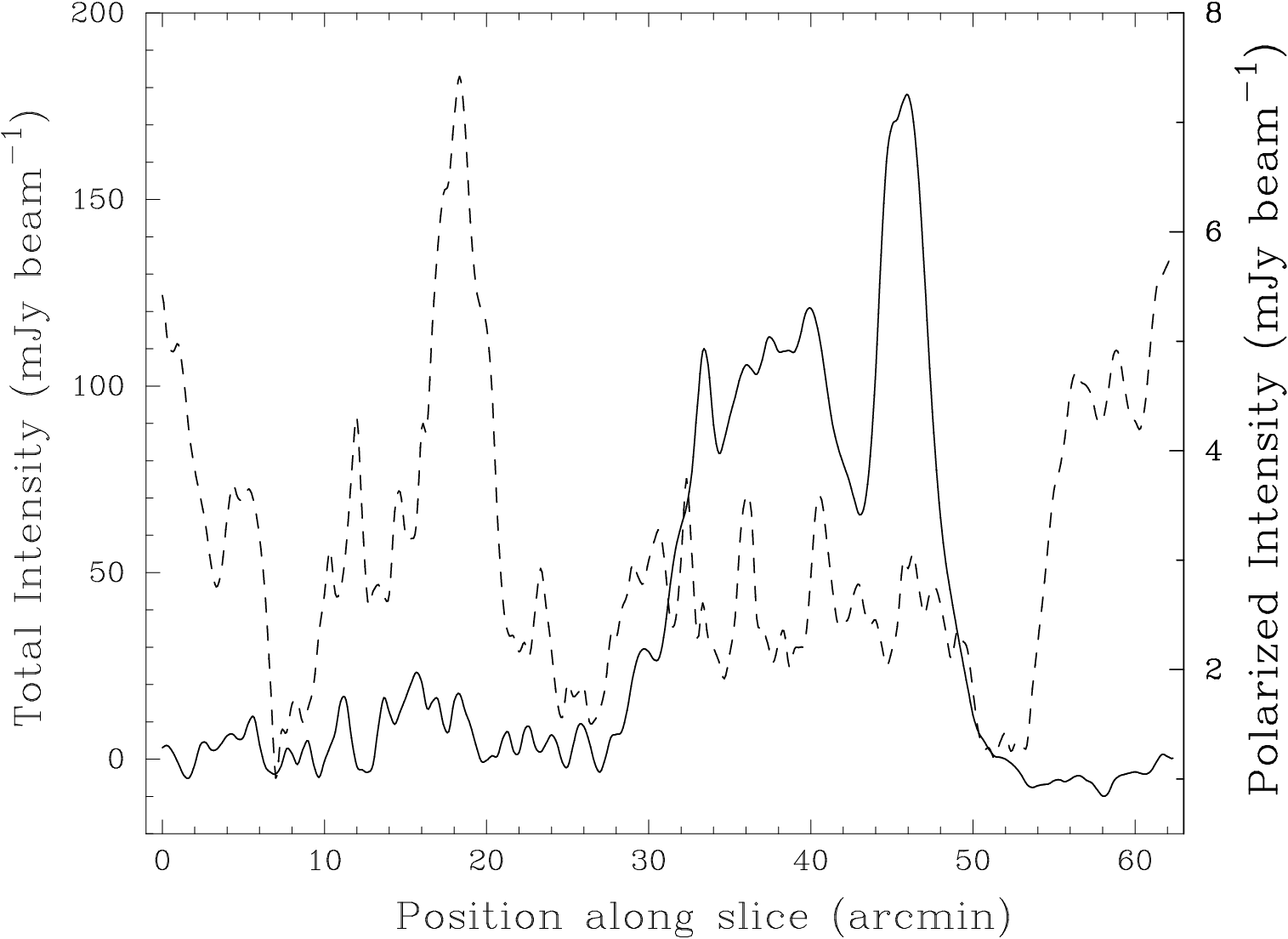,height=12cm,angle=270}}
\caption{Profiles of total intensity (solid line) and 
linearly polarized intensity (broken line) across RCW~94.
The slice along which these profiles was taken is
shown in Fig~\ref{fig_hii_pol}.}
\label{fig_rcw94_slice}
\end{figure}

\begin{figure}[htb]
\vspace{1.5cm}
\caption{Polarization properties of ``canals'' in a small
part of the Test Region. The greyscale represents
polarized position angle, while the contours
(at levels of 3, 3.5 and 4~mJy~beam$^{-1}$)
indicate polarized intensity. The boxes
represent rotation measure as in Fig~\ref{fig_ext_rm};
here the largest boxes shown correspond to RMs of
magnitude greater than $\pm30$~rad~m$^{-2}$. 
It can be seen that the position angle sharply changes
across each canal, but that there are no corresponding
changes in RM.}
\label{fig_canals}
\end{figure}

\begin{figure}[htb]
\centerline{\psfig{file=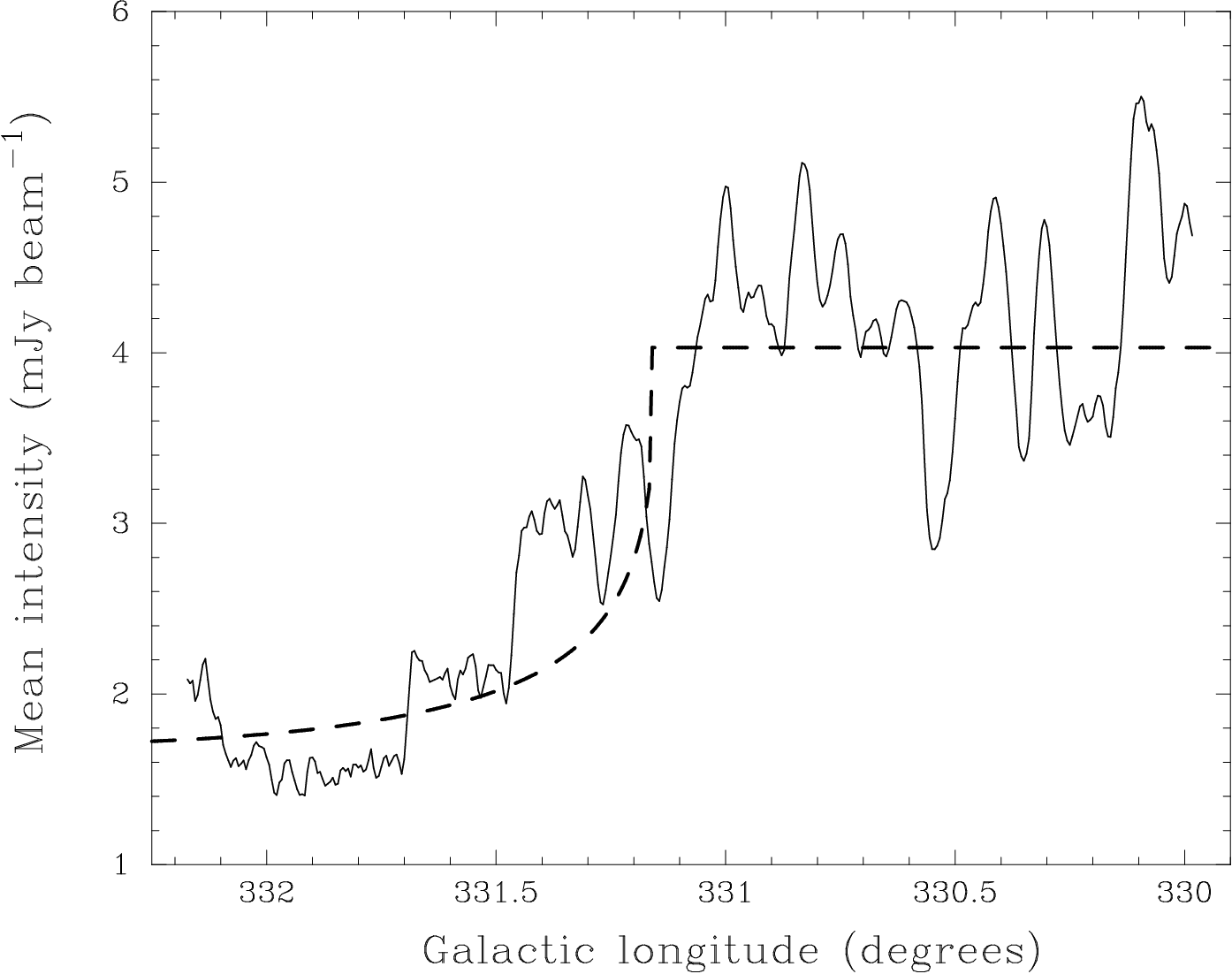,height=12cm,angle=270}}
\caption{Profile of polarized intensity as a function
of longitude across void 1, covering the region marked
with a box in Fig~\ref{fig_ext_L_2}.
The solid line represents the mean polarized intensity
of the data averaged over the range $0\fdg7 < b < 1\fdg2$.
The dashed curve is the depolarization predicted
by Equation~(\ref{eqn_beam_depol}), resulting
from RM fluctuations on scales of 1~arcmin with dispersion
$\sigma_{\rm RM} = 37$~rad~m$^{-2}$.}
\label{fig_void_slice}
\end{figure}

\begin{figure}[htb]
\centerline{\psfig{file=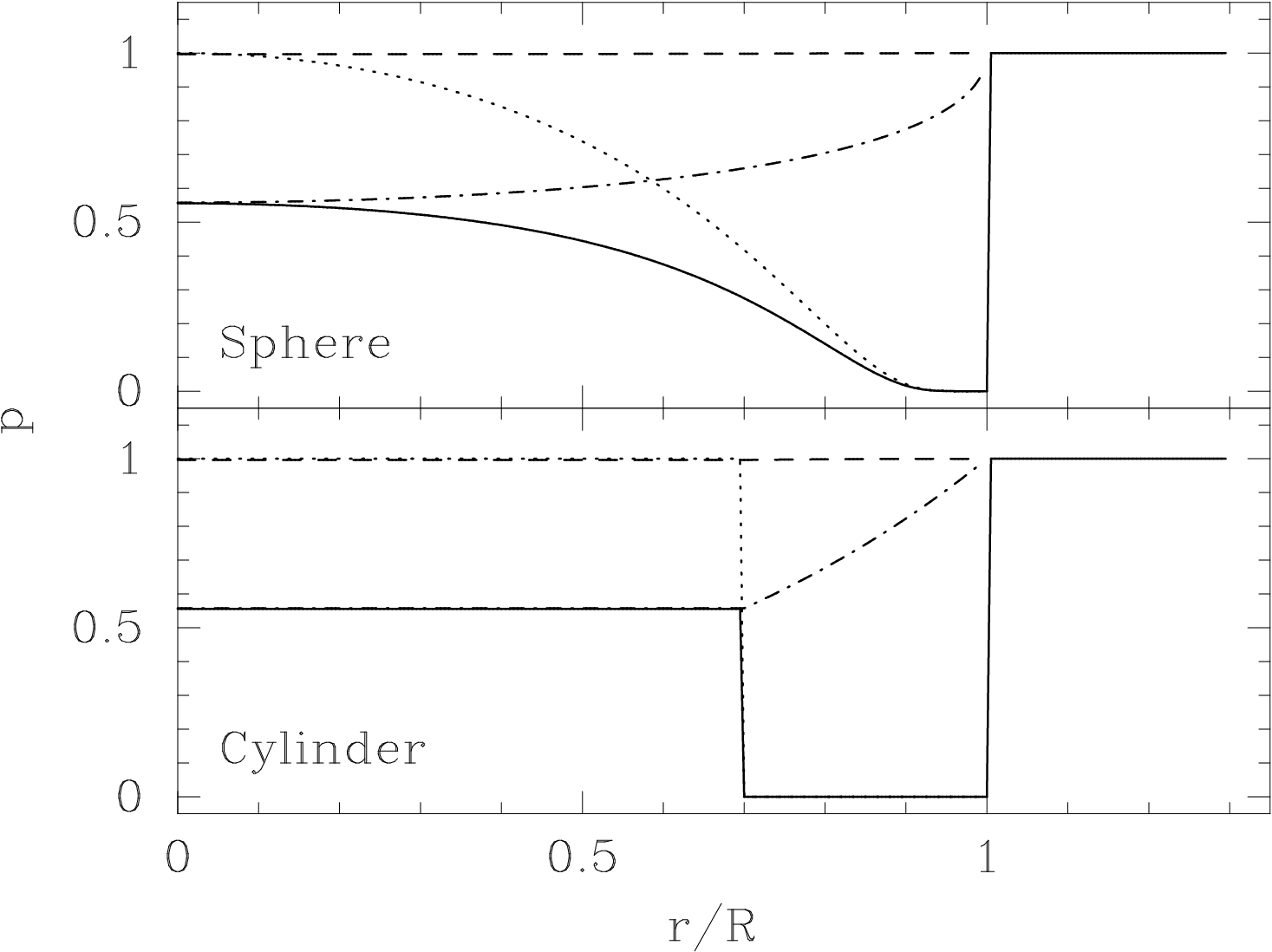,width=16cm,angle=270}}
\caption{Predicted depolarization factor as a function of distance
from the center of RCW~94, for spherical (upper) and cylindrical
(lower) geometries. In both cases the parameters assumed
were $2R = 28$~pc,
$n_e = 20$~cm$^{-3}$, $B_u = B_r = 1.2$~$\mu$G, $l = 0.15$~pc
and $\theta = 60^\circ$. The dashed,
dotted and dot-dashed lines correspond respectively
to bandwidth ($p_b$; Equation~[\ref{eqn_depol_band}]), RM-gradient ($p_g$; 
Equation~[\ref{eqn_depol_grad}])
and beam depolarization ($p_f$; Equation~[\ref{eqn_beam_depol}]).
The solid line represents the total depolarization ($p$; 
Equation~[\ref{eqn_depol_tot}]).}
\label{fig_rcw_sphere}
\end{figure}

\clearpage

\appendix

\section{Effect of an interferometer on measured rotation measure}
\label{appendix_1}

Suppose that there is a background source of emission,
with uniformly polarized intensity $P_0$ and uniform polarization
position angle $\Theta_0$. 
The complex polarization corresponding to such a source is:
\begin{equation}
{\bf P_0} = P_0~e^{2i\Theta_0}, 
\end{equation}
so that $Q_0 = P_0 \cos 2\Theta_0$
and $U_0 = P_0 \sin 2\Theta_0$. This emission passes
through a uniform screen of rotation measure $R_s$, and
then through a compact cloud of rotation measure $R_c$.

After passing through the uniform screen, the resultant complex
polarization is:
\begin{equation}
{\bf P_s} = {\bf P_0}~e^{2iR_s\lambda^2} = P_0~e^{2i(\Theta_0 +
R_s\lambda^2)}
\end{equation}

Along lines-of-sight which then pass through the cloud,
the complex polarization is:
\begin{equation}
{\bf P_c} = {\bf P_s}~e^{2iR_c\lambda^2}
= P_0~e^{2i(\Theta_0 + R_s\lambda^2 + R_c\lambda^2)}
\end{equation}

Suppose that this radiation field is then observed with an interferometer.
The interferometer is only sensitive to a limited range of
spatial scales (which we assume includes that corresponding
to the extent of the compact cloud), but has no sensitivity
to the uniform component of the emission. Thus the emission
detected by the instrument is the incident field, less
the uniform component.  

Thus on lines-of-sight which do not pass through the cloud,
the detected complex polarization is:
\begin{equation}
{\bf P_{det}} = {\bf P_s} - {\bf P_s} = 0,
\end{equation}
indicating that no emission is detected off-source.
However, towards the cloud, one detects:
\begin{equation}
{\bf P_{det}} = {\bf P_c} - {\bf P_s} = P_0~e^{2i\left(\Theta_0 + R_s\lambda^2
\right)} 
\left( e^{2iR_c\lambda^2} - 1 \right).
\end{equation}
This simplifies to:
\begin{equation}
{\bf P_{det}} = 2 P_0 \sin \left(R_c \lambda^2 \right) 
e^{2i\left(\Theta_0 + \frac{\pi}{4}  +
\lambda^2 \left[R_s + \frac{R_c}{2} \right] \right)},
\end{equation}
so that we can write:
\begin{equation}
{\bf P_{det}} = P_{\rm det}~e^{2i\left(\Theta_{\rm det} + {\rm RM}_{\rm det} 
\lambda^2\right)},
\label{eqn_appendix_final}
\end{equation}
where $P_{\rm det} = 2P_0 \sin\left(R_c \lambda^2 \right)$, 
$\Theta_{\rm det} = \Theta_0 + \frac{\pi}{4}$ and ${\rm RM}_{\rm det} = R_s +
\frac{R_c}{2}$.

Thus when observed with an interferometer, no polarized emission is
seen on lines-of-sight which do not pass through the cloud, since this
uniform component of the polarized emission is completely resolved out
by the interferometer.  Towards the cloud, polarized emission is
detected, but whose apparent polarized intensity can be as large as
twice that of the emitting source. The RM measured towards this
emission is the sum of the RM from the diffuse component and {\em
half}\ the RM of the cloud. After correction for Faraday rotation, the
intrinsic position angle of incident radiation inferred from the
observations is $45^\circ$ greater than the true position angle of the
emitted radiation.

\end{document}